\documentclass[aps, preprint]{revtex4-1}
\usepackage{amsmath}
\usepackage{amssymb}
\usepackage{fullpage}
\usepackage{graphicx}
\usepackage{xcolor}
\usepackage{afterpage}
\usepackage{mathtools}
\usepackage{color}
\usepackage{amsfonts}
\usepackage[sort&compress]{natbib}
\usepackage{xr}
\makeatletter
\newcommand{\RNum}[1]{\uppercase\expandafter{\romannumeral #1\relax}}
\newcommand{\vast}{\bBigg@{4}}
\newcommand{\Vast}{\bBigg@{5}}
\makeatother
\begin{document}
\title{The Role of Thermalizing and Non-thermalizing Walls in Phonon Heat Conduction along Thin Films}
\author{Navaneetha K. Ravichandran}
\author{Austin J. Minnich}
\email{aminnich@caltech.edu}
\affiliation{Division of Engineering and Applied Science,\\ California Institute of Technology, Pasadena, California 91125, USA}
\date{\today}
\begin{abstract}
Phonon boundary scattering is typically treated using the Fuchs-Sondheimer theory, which assumes that phonons are thermalized to the local temperature at the boundary. However, whether such a thermalization process actually occurs and its effect on thermal transport remains unclear. Here we examine thermal transport along thin films with both thermalizing and non-thermalizing walls by solving the spectral Boltzmann transport equation (BTE) for steady state and transient transport. We find that in steady state, the thermal transport is governed by the Fuchs-Sondheimer theory and is insensitive to whether the boundaries are thermalizing or not. In contrast, under transient conditions, the thermal decay rates are significantly different for thermalizing and non-thermalizing walls. We also show that, for transient transport, the thermalizing boundary condition is unphysical due to violation of heat flux conservation at the boundaries. Our results provide insights into the boundary scattering process of thermal phonons over a range of heating length scales that are useful for interpreting thermal measurements on nanostructures.
\end{abstract}
\maketitle
\section{Introduction}
Engineering the thermal conductivity of nanoscale materials has been a topic of considerable research interest over the past two decades~\cite{cahill_nanoscale_2014}. While applications such as GaN transistors~\cite{cho_phonon_2014, yan_graphene_2012} and light emitting diodes (LEDs)~\cite{koh_heat-transport_2009} require high thermal conductivity substrates to dissipate heat, the performance of thermoelectric and thermal insulation devices can be significantly enhanced by reducing  their thermal conductivity~\cite{poudel_high-thermoelectric_2008, biswas_high-performance_2012}. In many of these applications, phonon boundary scattering is the dominant resistance to heat flow, making the detailed understanding of this process essential for advancing applications.\\

Phonon boundary scattering has been studied extensively both theoretically and experimentally. The thermal conductivity reduction due to boundary scattering of phonons is conventionally treated using the Fuchs-Sondheimer theory, which was first derived for electron boundary scattering independently by Fuchs~\cite{fuchs_conductivity_1938} and Reuter and Sondheimer~\cite{sondheimer_theory_1948} and was later extended to phonon boundary scattering in several works~\cite{chen_nanoscale_2005, cuffe_reconstructing_2015, turney_in-plane_2010}. Fuchs-Sondheimer theory is widely used to interpret experiments but makes an important assumption that the diffusely scattered part of the phonon spectrum at a partially specular wall is at a local thermal equilibrium with the wall - the thermalizing boundary condition. The thermalizing boundary condition is also a key assumption in the diffuse boundary scattering limit of Casimir\textsc{\char13}s theory~\cite{casimir_note_1938}.\\ 

Several computational works~\cite{turney_in-plane_2010, aksamija_anisotropy_2010, hao_frequency-dependent_2009, mazumder_monte_2001, ravichandran_coherent_2014} have studied the reduction in thermal conductivity  due to phonon boundary scattering in nanostructures by solving the phonon Boltzmann transport equation (BTE). These works have considered either thermalizing or non-thermalizing boundaries but have never compared the effect of these two different boundary conditions on the thermal conductivity of nanostructures. Several experimental works have also studied the reduction in thermal conductivity of nanomaterials such as nanowires ~\cite{li_thermal_2003, hippalgaonkar_fabrication_2010, chen_thermal_2008}, thin films~\cite{cuffe_reconstructing_2015, johnson_direct_2013, maznev_lifetime_2015} and nanopatterned structures~\cite{tang_holey_2010} due to phonon boundary scattering. These works have used the Fuchs-Sondheimer theory to interpret their measurements. However, it is not clear if the assumptions made in the Fuchs-Sondheimer theory are necessarily applicable for these experiments. In fact, an analysis of the effect of the key assumption made in the Fuchs-Sondheimer theory, that the walls are thermalizing, has never been investigated due to the challenges involved in solving the BTE for non-thermalizing walls.\\
 
Here, we examine the role of thermalizing and non-thermalizing walls in heat conduction along thin films by solving the spectral phonon Boltzmann transport equation (BTE) for a suspended thin film under steady state and transient transport conditions. We find that steady state transport is insensitive to whether phonons are thermalized or not at the boundaries and that Fuchs-Sondheimer theory accurately describes thermal transport along the thin film. In the case of transient transport, we find that the decay rates are significantly different for thermalizing and non-thermalizing walls and that Fuchs-Sondheimer theory accurately predicts the thermal conductivity only when the thermal transport is diffusive. Moreover, under transient transport conditions, we find that phonons cannot undergo thermalization at the boundaries in general due to the violation of heat flux conservation. Our results provide insights into the boundary scattering process of thermal phonons that are useful for interpreting thermal measurements on nanostructures.\\
\section{Modeling}
\subsection{ Boltzmann Transport Equation} \label{general_bte}
We begin our analysis by considering the two dimensional spectral transient Boltzmann transport equation (BTE) under the relaxation time approximation for an isotropic crystal, given by,
\begin{equation}
\frac{\partial g_\omega}{\partial t} + \mu v_g\frac{\partial g_\omega}{\partial z} + v_g\sqrt{1-\mu^2}\cos\phi\frac{\partial g_\omega}{\partial x} = -\frac{g_\omega - g_o\left(T\right)}{\tau_\omega} + \frac{Q_\omega}{4\pi} \label{PBTE}
\end{equation} 
Here, $g_\omega$ is the phonon energy distribution function, $\omega$ is the phonon frequency, $v_g$ is the phonon group velocity, $\tau_\omega$ is the phonon relaxation time, $x$ and $z$ are the spatial coordinates, $t$ is the time variable, $g_0\left(T\right)$ is the equilibrium phonon distribution function at a deviational temperature $T = T_0 + \Delta T$ from an equilibrium temperature $T_0$, $\mu$ is the direction cosine, $\phi$ is the azimuthal angle and $Q_\omega$ is the rate of volumetric heat generation for each phonon mode. As the in-plane (x) direction is infinite in extent, we require boundary conditions only for the cross-plane ($z$) direction. In the traditional Fuchs-Sondheimer problem, the boundary conditions enforce that the diffusely scattered phonons are thermalized while also allowing some phonons to be specularly reflected. Here, we generalize these boundary conditions to allow for the possibility of both partial thermalization and partial specularity as:
\begin{align}
\label{bdry_conditions}
\begin{split}
\mathrm{For}\ \mu\in\left(0, 1\right], & \\
g^+_\omega\left(0, \mu, \phi\right) &= p_\omega g^-_\omega\left(0, -\mu, \phi\right) \\
&\ \ \ \ \ \ \ + \left(1- p_\omega \right)\Bigg( \sigma_\omega \frac{C_\omega\Delta T\left(z=0\right)}{4\pi} - \frac{\left(1- \sigma_\omega \right)}{\pi}\int_0^{2\pi}\int_{-1}^0g^-_\omega\left(0, \mu ', \phi'\right)\mu '\mathrm{d}\mu '\mathrm{d}\phi'\Bigg)\\
\mathrm{For}\ \mu\in\left[-1, 0\right), & \\
g^-_\omega\left(d, \mu, \phi\right) &= p_\omega g^+_\omega\left(d, -\mu, \phi\right) \\
&\ \ \ \ \ \ \ + \left(1- p_\omega \right)\Bigg( \sigma_\omega \frac{C_\omega\Delta T\left(z=d\right)}{4\pi} + \frac{\left(1- \sigma_\omega \right)}{\pi}\int_0^{2\pi}\int_0^1g^+_\omega\left(d, \mu ', \phi'\right)\mu '\mathrm{d}\mu '\mathrm{d}\phi'\Bigg)
\end{split}
\end{align}
where, $d$ is the thickness in the cross-plane direction, $g^+_\omega\left(0, \mu, \phi\right)$ is the phonon distribution leaving the cross-plane wall at $z=0$, $g^-_\omega\left(0, \mu, \phi\right)$ is the phonon distribution approaching the cross-plane wall at $z=0$, $g^+_\omega\left(d, \mu, \phi\right)$ is the phonon distribution approaching the cross-plane wall at $z=d$, $g^-_\omega\left(d, \mu, \phi\right)$ is the phonon distribution leaving the cross-plane wall at $z=d$, $C_\omega$ is the specific heat of a phonon mode with frequency $\omega$, $ p_\omega $ and $\sigma_\omega$ are the phonon specularity parameter and the thermalization parameter for the thin film walls respectively. The specularity parameter represents the fraction of specularly scattered phonons at the boundaries and the thermalization parameter represents the fraction of the phonon distribution that is absorbed and reemitted at the local equilibrium temperature of the thin film walls. For simplicity, we ignore mode conversion for non-thermalizing boundary condition in our analysis. \\

The unknown quantities in this problem are the phonon distribution function ($g_\omega\left(t, x, z, \mu, \phi\right)$) and the deviational temperature distribution ($\Delta T\left(t, x, z\right)$). They are related to each other through the energy conservation requirement,
\begin{equation}
\int_{\omega = 0}^{\omega_m}\int_{\mu = -1}^1\int_{\phi = 0}^{2\pi}\left[\frac{g _\omega}{\tau_\omega} - \frac{1}{4\pi}\frac{C_\omega}{\tau_\omega}\Delta T\right]\mathrm{d}\phi\mathrm{d}\mu\mathrm{d}\omega = 0 \label{e_conserve}
\end{equation}
Due to the high dimensionality of the BTE, analytical or semi-analytical solutions are only available in literature for either semi-infinite domains~\cite{minnich_multidimensional_2015, hua_analytical_2014, hua_transport_2014} or domains with simple boundary and transport conditions~\cite{hua_semi-analytical_2015} or with several approximations~\cite{zeng_phonon_2000}. For nanostructures with physically realistic boundaries, several numerical solutions of the BTE have been reported ~\cite{turney_in-plane_2010, peraud_monte_2014, ravichandran_coherent_2014}. However, computationally efficient analytical or semi-analytical solutions for the in-plane heat conduction along even simple unpatterned  films~\cite{johnson_direct_2013, cuffe_reconstructing_2015} are unavailable. To overcome this problem, we solve the BTE analytically for steady state transport (section~\ref{steady_state}) and semi-analytically for transient transport along thin films in the TG experiment~\cite{johnson_direct_2013, cuffe_reconstructing_2015} (section~\ref{transient}).

\subsection{Steady State Heat Conduction in Thin Films} \label{steady_state}
In this section, we extend the Fuchs-Sondheimer relation for thermal conductivity suppression due to phonon boundary scattering to the general boundary conditions described in equation~\ref{bdry_conditions}. To simulate steady state transport, $Q_\omega$ is set to 0 in the BTE (equation~\ref{PBTE}). Furthermore, we assume that a one-dimensional temperature gradient exists along the thin film and $\frac{\partial g_\omega}{\partial x}\approx\frac{\partial g^0_\omega}{\partial x}$. These assumptions are consistent with the conditions under which typical steady state thermal transport measurements are conducted on nanostructures~\cite{liu_phononboundary_2004, chen_thermal_2008, ju_phonon_1999}. Under these assumptions, the BTE is simplified as,
\begin{equation}
v_g\mu\frac{\partial g_\omega}{\partial z} + v_g\sqrt{1-\mu^2}\cos\phi\frac{\partial g^0_\omega}{\partial x} = -\frac{g_\omega - g_\omega^0}{\tau_\omega} \label{PBTE_steady}
\end{equation}
For steady state transport, it is convenient to solve the BTE in terms of the deviation from equilibrium distribution ($\bar{g}_\omega = g_\omega - g_\omega^0\left(\Delta T\left(x\right)\right)$). In this case, the BTE transforms into,
\begin{equation}
\frac{\partial \bar{g}_\omega}{\partial z} + \frac{\bar{g}_\omega}{\mu\Lambda_\omega} = -\frac{\cos\phi\sqrt{1-\mu^2}}{\mu}\frac{\partial g^0_\omega}{\partial x}	\label{PBTE_steady_dev}
\end{equation}
The boundary conditions (equation~\ref{bdry_conditions}) for $\bar{g}_\omega$ now become,
\begin{align}
\label{Steady_gen_BCs_1}
\begin{split}
\mathrm{For}\ \mu\in\left(0, 1\right]&,\\
\bar{g}_\omega ^+\left(0, \mu, \phi\right) &= p_\omega \bar{g}_\omega ^-\left(0, -\mu, \phi\right) - \frac{\left(1- p_\omega \right)\left(1- \sigma_\omega \right)}{\pi}\int_0^{2\pi}\int_{-1}^0\bar{g}_\omega^-\left(0, \mu', \phi\right)\mu'\mathrm{d}\mu'\mathrm{d}\phi\\
\mathrm{For}\ \mu\in\left[-1, 0\right)&,\\
\bar{g}_\omega ^-\left(d, \mu, \phi\right) &= p_\omega \bar{g}_\omega ^+\left(d, -\mu, \phi\right) + \frac{\left(1- p_\omega \right)\left(1- \sigma_\omega \right)}{\pi}\int_0^{2\pi}\int_0^1\bar{g}_\omega^+\left(d, \mu', \phi\right)\mu'\mathrm{d}\mu'\mathrm{d}\phi
\end{split}
\end{align}
The general solution of the BTE (equation~\ref{PBTE_steady_dev}) along with the boundary conditions (equation~\ref{Steady_gen_BCs_1}) is given by,
\begin{align}
\label{Steady_gen_soln_2}
\begin{split}
\bar{g}_\omega^+\left(z, \mu, \phi\right) &= -\Lambda_\omega\cos\phi\sqrt{1-\mu^2}\frac{\partial g_\omega^0}{\partial x}\vast[1-\frac{\exp\left(-\frac{z}{\mu\Lambda_\omega}\right)\left(1- p_\omega \right) }{1- p_\omega \exp\left(-\frac{d}{\mu\Lambda_\omega}\right)}\vast] \\
&\ \ \ \ \ + \underbrace{\frac{\left(1- p_\omega \right)\left(1- \sigma_\omega \right)\left[A^+_\omega + p_\omega \exp\left(-\frac{d}{\mu\Lambda_\omega}\right)A^-_\omega\right]}{1 - p_\omega^2\exp\left(-\frac{2d}{\mu\Lambda_\omega}\right)}\exp\left(-\frac{z}{\mu\Lambda_\omega}\right)}_{\mathrm{\RNum{1}}}\\
\bar{g}_\omega^-\left(z, -\mu, \phi\right) &= -\Lambda_\omega\cos\phi\sqrt{1-\mu^2}\frac{\partial g_\omega^0}{\partial x}\vast[1-\frac{\exp\left(-\frac{\left(d-z\right)}{\mu\Lambda_\omega}\right)\left(1- p_\omega \right)}{1- p_\omega \exp\left(-\frac{d}{\mu\Lambda_\omega}\right)}\vast] \\
&\ \ \ \ \ + \underbrace{\frac{\left(1- p_\omega \right)\left(1- \sigma_\omega \right)\left[A^-_\omega + p_\omega \exp\left(-\frac{d}{\mu\Lambda_\omega}\right)A^+_\omega\right]}{1 - p_\omega^2 \exp\left(-\frac{2d}{\mu\Lambda_\omega}\right)}\exp\left(-\frac{\left(d-z\right)}{\mu\Lambda_\omega}\right)}_{\mathrm{\RNum{2}}}\\
\end{split}
\end{align}
for $\mu\in\left(0, 1\right]$. Here, the terms $A^+_\omega$ and $A^-_\omega$ only depend on phonon frequency. In particular, they are independent of the angular coordinates $\mu$ and $\phi$. The derivation of the final expressions for $\bar{g}_\omega^+\left(z, \mu, \phi\right)$ and $\bar{g}_\omega^-\left(z, -\mu, \phi\right)$ (equation~\ref{Steady_gen_soln_2}) is shown in section I of the supplementary material. The expression for the in-plane ($x$ direction) spectral heat flux is given by,
\begin{align}
\label{x-heat_flux_1}
\begin{split}
q_{x, \omega} &= \frac{1}{d}\int_{z=0}^d\int_{\mu = -1}^1\int_{\phi=0}^{2\pi}v_x\bar{g}_\omega\frac{D\left(\omega\right)}{4\pi}\mathrm{d}\phi\mathrm{d}\mu\mathrm{d}z\\
&= -\Bigg[\frac{1}{3}C_\omega v_g\Lambda_\omega\Bigg]\frac{\partial T}{\partial x} \left[1 - \frac{3\left(1-p_\omega\right)\Lambda_\omega}{2d}\int_0^1\left(\mu-\mu^3\right)\frac{1-\exp\left(-\frac{d}{\mu\Lambda_\omega}\right)}{1-p_\omega\exp\left(-\frac{d}{\mu\Lambda_\omega}\right)}\mathrm{d}\mu\right]
\end{split}
\end{align}
since the diffuse contributions to the distribution functions $\bar{g}_\omega^+\left(z, \mu, \phi\right)$ and $\bar{g}_\omega^-\left(z, -\mu, \phi\right)$ (terms~\RNum{1} and~\RNum{2} in equation~\ref{Steady_gen_soln_2}) are independent of the azimuthal angle $\phi$ and integrate out to $0$. Comparing equation~\ref{x-heat_flux_1} with the expression for heat flux from the Fourier's law, the spectral effective thermal conductivity of the thin film is obtained as a product of the bulk spectral thermal conductivity and the well-known Fuchs-Sondheimer reduction factor due to phonon boundary scattering given by,
\begin{align}
\label{Steady_k_eff}
\begin{split}
k_{\omega, \mathrm{eff}}\left(d\right) &= \underbrace{\Bigg[\frac{1}{3}C_\omega v_g\Lambda_\omega\Bigg]}_{k_{\omega, \mathrm{bulk}}} \underbrace{\left[1 - \frac{3\left(1-p_\omega\right)\Lambda_\omega}{2d}\int_0^1\left(\mu-\mu^3\right)\frac{1-\exp\left(-\frac{d}{\mu\Lambda_\omega}\right)}{1-p_\omega\exp\left(-\frac{d}{\mu\Lambda_\omega}\right)}\mathrm{d}\mu\right]}_{\mathrm{Fuchs-Sondheimer\ reduction\ factor} - F\left(\frac{\Lambda_\omega}{d}\right)}
\end{split}
\end{align}
It is interesting to observe from equation~\ref{Steady_k_eff} that the spectral effective thermal conductivity is independent of the thermalization parameter $ \sigma_\omega $ even though a general boundary condition (equation~\ref{bdry_conditions}) has been used in this derivation. Thus, the steady state thermal conductivity suppression due to boundary scattering is only influenced by the relative extent of specular and diffuse scattering (parameterized by the specularity parameter $ p_\omega $) and does not depend on the type of diffuse scattering process (parameterized by the thermalization parameter $ \sigma_\omega $). We explicitly demonstrate this result using numerical simulations in section~\ref{Steady_Results}. \\
\subsection{Transient Heat Conduction in Thin Films} \label{transient}
In this section, we solve the BTE (equation~\ref{PBTE}) for transient thermal transport along a thin film. The initial temperature profile considered in this work is identical to that which occurs in the Transient Grating (TG) experiment, which has been used extensively to study heat conduction in suspended thin films~\cite{johnson_direct_2013, cuffe_reconstructing_2015}. In the TG experiment, the thermal transport properties of the sample are obtained by observing the transient decay of a one-dimensional impulsive sinusoidal temperature grating on the sample at different grating periods. In the large grating period limit of heat diffusion, the temporal decay is a single exponential. Since the initial temperature distribution is an infinite one-dimensional sinusoid in the $x$ direction, the temperature distribution remains spatially sinusoidal at all later times. Therefore, each wave vector $q$ in the spatially Fourier transformed BTE directly corresponds to a unique grating period $\lambda = 2\pi/q$. Unlike in the steady state case, here we solve for the absolute phonon distribution $g_\omega$ rather than the deviation $\bar{g}_\omega = g_\omega - g_\omega^0$. Furthermore, the BTE is solved in the frequency domain ($\eta$) by Fourier transforming equation~\ref{PBTE} in the time variable $t$. With these transformations, the BTE reduces to,
\begin{equation}
i\eta G _\omega + \mu v_g\frac{\partial G _\omega}{\partial z} + i q v_g\sqrt{1-\mu^2}\cos\phi\ G _\omega = -\frac{ G _\omega}{\tau_\omega} + \frac{1}{4\pi}\frac{C_\omega}{\tau_\omega}\Delta\bar{T} + \frac{\bar{Q}_\omega}{4\pi} \label{BTE_FT}
\end{equation}
where, the substitution $ G _0\left(T\right) = \frac{1}{4\pi}C_\omega\Delta \bar{T}$ has been made and $ G _\omega$ represents the spatial (in-plane axis) and temporal Fourier transform of absolute phonon energy distribution function $g_\omega$. \\

The outline of the solution methodology for equation~\ref{BTE_FT} is as follows. The general solution is given by,\\
\begin{align}
\label{Gen_soln}
\begin{split}
\mathrm{For}\ \mu\in\left(0, 1\right], \ \ \ G ^+_\omega\left(z, \mu, \phi\right) &= G ^+_\omega\left(0, \mu, \phi\right)\exp\left(-\frac{\gamma^{\mathrm{FS}}_{\mu\phi}}{\mu\Lambda_\omega}z\right) \\
&\ \ \ \ \ + \frac{\exp\left(-\frac{\gamma^{\mathrm{FS}}_{\mu\phi}}{\mu\Lambda_\omega}z\right)}{4\pi\mu\Lambda_\omega}\int_0^z\left(C_\omega\Delta\bar{T} + \bar{Q}_\omega\tau_\omega\right)\exp\left(\frac{\gamma^{\mathrm{FS}}_{\mu\phi}}{\mu\Lambda_\omega}z'\right)\mathrm{d}z'\\
\mathrm{For}\ \mu\in\left[-1, 0\right), \ \ \ 
G ^-_\omega\left(z, \mu, \phi\right) &= G ^-_\omega\left(d, \mu, \phi\right)\exp\left(\frac{\gamma^{\mathrm{FS}}_{\mu\phi}}{\mu\Lambda_\omega}\left(d-z\right)\right) \\
&\ \ \ \ \ - \frac{\exp\left(-\frac{\gamma^{\mathrm{FS}}_{\mu\phi}}{\mu\Lambda_\omega}z\right)}{4\pi\mu\Lambda_\omega}\int_z^d\left(C_\omega\Delta\bar{T} + \bar{Q}_\omega\tau_\omega\right)\exp\left(\frac{\gamma^{\mathrm{FS}}_{\mu\phi}}{\mu\Lambda_\omega}z'\right)\mathrm{d}z'\\
\mathrm{where},\ \ \ \ &\gamma^{\mathrm{FS}}_{\mu\phi} = \left(1 + i\eta\tau_\omega\right) + i\Lambda_\omega q \sqrt{1-\mu ^2}\cos\phi
\end{split}
\end{align}
Here, $ G ^+_\omega\left(0, \mu, \phi\right)$ and $ G ^-_\omega\left(d, \mu, \phi\right)$ are determined by solving the boundary conditions (equation~\ref{bdry_conditions}) with the following procedure. First, the angular integrals in the boundary conditions are discretized using Gauss quadrature, which results in the following set of linear equations in the variables $ G ^+_\omega\left(0, \mu_i, \phi_j\right)$ and $ G ^-_\omega\left(d, -\mu_i, \phi_j\right)$ for every $\{\mu_i, \phi_j\}\in\left(0, 1\right]\times\left[0, 2\pi\right]$ doublet from the discretization. 
\begin{align}
\label{bdry_gen_sub}
\begin{split}
G^+_\omega\left(0, \mu_i, \phi_j\right) &= p_\omega G^-_\omega\left(d, -\mu_i, \phi_j\right)\exp\left(-\frac{\gamma^{\mathrm{FS}}_{ij}}{\mu_i\Lambda_\omega}d\right) \\
&\ \ \ \ \ + \frac{ p_\omega }{4\pi\mu_i\Lambda_\omega}\int_0^d\left(C_\omega\Delta\bar{T} + \bar{\tilde{Q}}_\omega\tau_\omega\right)\exp\left(-\frac{\gamma^{\mathrm{FS}}_{ij}}{\mu_i\Lambda_\omega}z'\right)\mathrm{d}z'\\
&\ \ \ \ \ + \left(1- p_\omega \right)\Bigg[ \sigma_\omega \frac{C_\omega\Delta \bar{T}\left(z=0\right)}{4\pi} \\
&\ \ \ \ \ + \frac{\left(1- \sigma_\omega \right)}{\pi}\sum_{i'j'}G^-_\omega\left(d, -\mu'_i, \phi'_j\right)\exp\left(-\frac{\gamma^{\mathrm{FS}}_{i'j'}}{\mu '_i\Lambda_\omega}d\right)\mu'_iw_{\mu'_i}w_{\phi'_j}\\
&\ \ \ \ \ + \frac{\left(1- \sigma_\omega \right)}{4\pi^2\Lambda_\omega}\sum_{i'j'}\int_0^d\left(C_\omega\Delta\bar{T} + \bar{\tilde{Q}}_\omega\tau_\omega\right)\exp\left(-\frac{\gamma^{\mathrm{FS}}_{i'j'}}{\mu '_i\Lambda_\omega}z'\right)\mathrm{d}z'w_{\mu'_i}w_{\phi'_j}\Bigg]\\
G^-_\omega\left(d, -\mu_i, \phi_j\right) &= p_\omega G^+_\omega\left(0, \mu_i, \phi_j\right)\exp\left(-\frac{\gamma^{\mathrm{FS}}_{ij}}{\mu_i\Lambda_\omega}d\right) \\
&\ \ \ \ \ + \frac{ p_\omega }{4\pi\mu_i\Lambda_\omega}\int_0^d\left(C_\omega\Delta\bar{T} + \bar{\tilde{Q}}_\omega\tau_\omega\right)\exp\left(-\frac{\gamma^{\mathrm{FS}}_{ij}}{\mu_i\Lambda_\omega}\left(d-z'\right)\right)\mathrm{d}z'\\
&\ \ \ \ \ + \left(1- p_\omega \right)\Bigg[ \sigma_\omega \frac{C_\omega\Delta \bar{T}\left(z=d\right)}{4\pi} \\
&\ \ \ \ \ + \frac{\left(1- \sigma_\omega \right)}{\pi}\sum_{i'j'}G^+_\omega\left(0, \mu'_i, \phi'_j\right)\exp\left(-\frac{\gamma^{\mathrm{FS}}_{i'j'}}{\mu '_i\Lambda_\omega}d\right)\mu'_iw_{\mu'_i}w_{\phi'_j}\\
&\ \ \ \ \ + \frac{\left(1- \sigma_\omega \right)}{4\pi^2\Lambda_\omega}\sum_{i'j'}\int_0^d\left(C_\omega\Delta\bar{T} + \bar{\tilde{Q}}_\omega\tau_\omega\right)\exp\left(-\frac{\gamma^{\mathrm{FS}}_{i'j'}}{\mu'_i\Lambda_\omega}\left(d-z'\right)\right)\mathrm{d}z'w_{\mu'_i}w_{\phi'_j}\Bigg]\\
\end{split}
\end{align}
To obtain equation~\ref{bdry_gen_sub}, we have substituted the general BTE solution into the boundary conditions to eliminate $ G ^-_\omega\left(0, \mu, \phi\right)$ and $ G ^+_\omega\left(d, \mu, \phi\right)$. Therefore, the only unknowns in the set of linear equations (equation~\ref{bdry_gen_sub}) are $ G ^+_\omega\left(0, \mu, \phi\right)$ and $ G ^-_\omega\left(d, \mu, \phi\right)$. By bringing the terms containing $ G ^+_\omega\left(0, \mu, \phi\right)$ and $ G ^-_\omega\left(d, \mu, \phi\right)$ to the left hand side, equation~\ref{bdry_gen_sub} can be written in a concise matrix form:
\begin{equation}
\left[\begin{array}{cc}
U^{+}_{kk'} & U^{-}_{kk'}\\
D^{+}_{kk'} & D^{-}_{kk'}
\end{array}\right]\left(\begin{array}{c}
G ^+_\omega\left(0, \mu_i, \phi_j\right)\\
G ^-_\omega\left(d, -\mu_i, \phi_j\right)
\end{array}\right) = \left( \begin{array}{c}
\bar{\tilde{c}}^+_\omega\left(0, \mu_{i'}, \phi_{j'}\right)\\
\bar{\tilde{c}}^-_\omega\left(d, \mu_{i'}, \phi_{j'}\right)
\end{array}\right) \label{bdry_soln_maintxt}
\end{equation}
where, $\bar{\tilde{c}}^+_\omega\left(0, \mu_{i'}, \phi_{j'}\right)$ and $\bar{\tilde{c}}^-_\omega\left(d, \mu_{i'}, \phi_{j'}\right)$ are analytical functions of the unknown temperature distribution function $\Delta\bar{T}$ obtained from the right hand side of equation~\ref{bdry_gen_sub}. The solution to this set of linear equations can be represented as:
\begin{equation}
\left(\begin{array}{c}
G ^+_\omega\left(0, \mu_i, \phi_j\right)\\
G ^-_\omega\left(d, -\mu_i, \phi_j\right)
\end{array}\right) = \left[\begin{array}{cc}
T^{+}_{kk'} & T^{-}_{kk'}\\
B^{+}_{kk'} & B^{-}_{kk'}
\end{array}\right]\left( \begin{array}{c}
\bar{\tilde{c}}^+_\omega\left(0, \mu_{i'}, \phi_{j'}\right)\\
\bar{\tilde{c}}^-_\omega\left(d, \mu_{i'}, \phi_{j'}\right)
\end{array}\right) \label{bdry_soln_maintxt}
\end{equation}
where $k$ is the index which represents the doublet $\{\mu_i, \phi_j\}$. The details of the simplification of the boundary conditions and the evaluation of $T^{+}_{kk'}$, $T^{-}_{kk'}$, $B^{+}_{kk'}$, $B^{-}_{kk'}$, $\bar{\tilde{c}}^+_\omega\left(0, \mu_{i'}, \phi_{j'}\right)$ and $\bar{\tilde{c}}^-_\omega\left(d, \mu_{i'}, \phi_{j'}\right)$ are described in section II A of the supplementary material. To close the problem, the expressions for $ G ^+_\omega\left(z, \mu, \phi\right)$ and $ G ^-_\omega\left(z, \mu, \phi\right)$ (equation~\ref{Gen_soln}) and the boundary conditions (equation~\ref{bdry_soln_maintxt}) are substituted into the energy conservation equation (equation~\ref{e_conserve}) and an integral equation in the variable $z$ for $\Delta \bar{T}\left(z\right)$ at each $\eta$ and $ q $ is obtained, which has the form:
\begin{equation}
\Delta\bar{T}\left(z\right) = h\left(z\right) + f\left(z\right) + \int_0^d\left[K\left(z', z\right)\Delta\bar{T}\left(z'\right)\right]\mathrm{d}z' \label{fredholm_2nd_maintxt}
\end{equation}
where the functional form of the inhomogeneous parts $f\left(z\right)$, $h\left(z\right)$ and the kernel $K\left(z', z\right)$ are described in section II B of the supplementary material. This integral equation (equation~\ref{fredholm_2nd_maintxt}) is then solved using the method of degenerate kernels for each $\eta$ and $ q $ to obtain the frequency domain solution $\Delta \bar{T}\left(z\right)$ for every $\eta$ and $ q $. The details of the degenerate kernel calculations are described in section II C of the supplementary material. Finally, the solution $\Delta \bar{T}\left(z\right)$ is substituted into equation~\ref{Gen_soln} to obtain expressions for $ G_\omega\left(z, \mu, \phi\right)$ and also the thickness-averaged in-plane heat flux $j_{x, \omega}$ given by,
\begin{align}
\label{Heat_flux}
\begin{split}
j_{x, \omega} &= \frac{1}{4\pi d}\int_0^d\int_0^{2\pi}\int_{-1}^1 G _\omega v_g\sqrt{1-\mu ^2}\cos\phi\mathrm{d}\mu\mathrm{d}\phi\mathrm{d}z \\
&=\frac{1}{4\pi}\sum_{ij}\vast[\frac{\mu_i\mathrm{Kn}^d_\omega}{\gamma^{\mathrm{FS}}_{ij}}\sum_{i'j'}\Bigg[\left(T^{+}_{kk'} + B^{+}_{kk'}\right)\bar{\tilde{c}}^+_\omega\left(0, \mu_{i'}, \phi_{j'}\right) \\
&\ \ \ \ \ + \left(T^{-}_{kk'} + B^{-}_{kk'}\right)\bar{\tilde{c}}^-_\omega\left(d, \mu_{i'}, \phi_{j'}\right)\Bigg]\left(1-\exp\left(-\frac{\gamma^{\mathrm{FS}}_{ij}}{\mu_i\mathrm{Kn}^d_\omega}\right)\right)\\
&\ \ \ \ \ + \frac{2}{4\pi\gamma^{\mathrm{FS}}_{ij}}\left(C_\omega\frac{t_0}{2} + \bar{Q}_\omega\tau_\omega\right)\left[1-\frac{\mu_i\mathrm{Kn}^d_\omega}{\gamma^{\mathrm{FS}}_{ij}}\left(1-\exp\left(-\frac{\gamma^{\mathrm{FS}}_{ij}}{\mu_i\mathrm{Kn}^d_\omega}\right)\right)\right] \\
&\ \ \ \ \ - \frac{C_\omega\left(1-\exp\left(-\frac{\gamma^{\mathrm{FS}}_{ij}}{\mu_i\mathrm{Kn}^d_\omega}\right)\right)}{4\pi\mu_i\mathrm{Kn}^d_\omega}\sum_{m=1}^Nt_m\frac{1+\left(-1\right)^m}{m^2\pi^2 + \left(\frac{\gamma^{\mathrm{FS}}_{ij}}{\mu_i\mathrm{Kn}^d_\omega}\right)^2}\vast]v_g\sqrt{1-\mu_i^2}w_{\mu_i}w_{\phi_j}\\
\end{split}
\end{align}
where $t_i$'s are the Fourier coefficients for the expansion of $\Delta\bar{T}$ in the cross-plane ($z$) direction and $\mathrm{Kn}^d_\omega = \Lambda_\omega/d$ is the Knudsen number. The conventional approach to describe the thermal transport properties of the thin film is to compare the expression for heat flux from the BTE solution with that expected for heat diffusion, as was done in equation~\ref{x-heat_flux_1} for the steady state Fuchs-Sondheimer theory. However, in practice, equation~\ref{Heat_flux} is not easily reduced into the form of Fourier's law. To overcome this problem, the following strategy is adopted. The solution of the Fourier heat equation to a one-dimensional heat conduction with an instantaneous spatially sinusoidal heat source is a simple exponential decay $\Delta T\left(t, x=0\right) = \Delta T_0\exp\left(-\gamma t\right)$, where the decay rate ($\gamma$) is related to the effective thermal conductivity ($k_{\mathrm{eff}}$) and the volumetric heat capacity of the solid ($C$) as, $\gamma = k_{\mathrm{eff}}q^2/C$. Therefore, to obtain the effective thermal conductivity from our calculations, we perform an inverse Fourier transform of the temperature distribution averaged in the z-direction ($\int_0^d\Delta\bar{T}\left(\eta, q, z\right)\mathrm{d}z$) with respect to the variable $\eta$, fit the resulting solution to an exponentially decaying function $\Delta T_0\exp\left(-\gamma t\right)$ and extract the thermal conductivity from the fit. If the fitting fails, the transport is in the strongly quasi-ballistic regime~\cite{hua_transport_2014} and we conclude that the Fourier law description of the heat conduction with an effective thermal conductivity $k_{\mathrm{eff}}$ is not valid for that case.\\

The semi-analytical solution of the BTE for transient transport presented in this work is computationally very efficient, taking only a few seconds on a single computer processor, while the direct Monte Carlo simulation of the BTE takes up to a few days on a high-performance computer cluster executed in parallel mode. Moreover, it is computationally challenging to extract the heat flux distribution directly from the Monte Carlo solution, while in our semi-analytical solution, the evaluation of heat flux distribution is a single step process (equation~\ref{Heat_flux}).\\

\section{Results \& Discussion}	\label{results}
We now present the results of the calculations for free-standing silicon thin films. To obtain these results, we use an isotropic dispersion and intrinsic scattering rates calculated using a Gaussian kernel-based regression~\cite{mingo_length-scale_2014} from the \textsl{ab-initio} phonon properties of isotopically pure silicon. The first principles phonon properties are calculated by \textsl{J. Carette \& N. Mingo} using ShengBTE~\cite{li_thermal_2012, li_shengbte:_2014} and Phonopy~\cite{togo_phonopy_2015} from the inter-atomic force constants calculated using VASP~\cite{kresse_textitab_1993, kresse_textitab_1994, kresse_efficiency_1996, kresse_efficient_1996}. 

\subsection{Steady State Transport in Thin Films}	\label{Steady_Results}
\subsubsection{Comparison with Monte Carlo Solution}	\label{MC_Steady}
\begin{figure}[!ht]
\begin{center}
\includegraphics*[scale=0.585, trim={100mm 90mm 90mm 90mm}, clip]{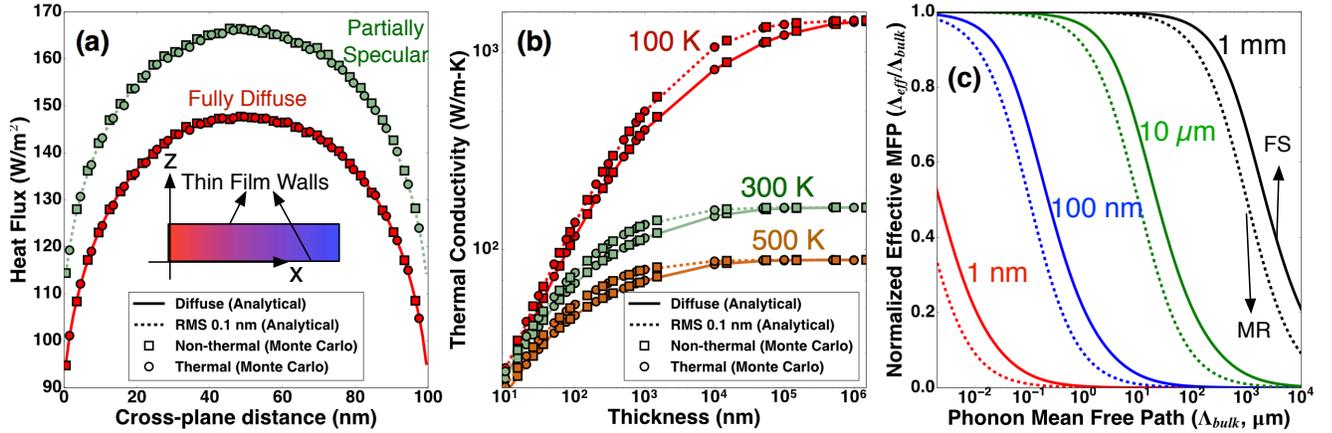}
\end{center}
\caption{(a) Comparison of the cross-plane distribution of the in-plane steady state heat flux between analytical and Monte Carlo solutions of the BTE at 300 K and film thickness of 100 nm for different boundary conditions. The geometry of the thin film and the coordinate axes used in this work are shown in the inset. (b) Comparison of the steady state thermal conductivity between analytical and Monte Carlo solutions of the BTE at different temperatures and thin film thicknesses. For both (a) and (b), the Monte Carlo solutions are identical for thermalizing and non-thermalizing boundary scattering and agree well with the analytical solution derived in this work for both fully diffuse and partially specular boundary conditions (RMS 0.1 nm). For the partially specular boundary condition, the specularity parameter ($p_\omega$) is calculated from Ziman's specularity model~\cite{ziman_electrons_1960} for a surface RMS roughness of 0.1 nm. (c) Effective MFPs of phonons computed using the Matthiessen's rule (MR) and the Fuchs-Sondheimer (FS) theory for different film thicknesses and fully diffuse boundary scattering. Matthiessen's rule underpredicts the effective phonon MFPs in thin films compared to the Fuchs-Sondheimer theory, which is a rigorous BTE solution.} \label{fig:Steady}
\end{figure}

We first examine steady state heat condition along thin films. Figures ~\ref{fig:Steady} (a) and (b) show the cross-plane distribution of the in-plane heat flux and the effective thermal conductivity respectively, for steady state transport through thin films computed using a Monte Carlo technique and the analytical solution from this work. The details of the Monte Carlo technique used in this work is described in section III of the the supplementary material. For both fully diffuse and partially specular boundary conditions, the heat flux distribution and the effective thermal conductivity of the thin film show excellent agreement between the Monte Carlo solutions and the analytical solution from this work over a range of temperatures and film thicknesses. In particular, both solutions predict identical heat flux and thermal conductivities for thermalizing and non-thermalizing boundary conditions at the thin film walls since the steady state transport is insensitive to the type of diffuse boundary scattering of phonons, as discussed in section~\ref{steady_state}. This observation can be generalized further to state that in steady state thermal transport experiments on thin films, it is impossible to distinguish between non-thermalizing and \textsl{any} type of inelastic diffuse scattering of phonons at boundaries.
\subsubsection{Effective Phonon Mean Free Path}
We also examine the effective mean free path (MFP) of phonons within the thin film for various film thicknesses. An approach to estimate the effective phonon mean free path in thin films is by using the Matthiessen's rule~\cite{ziman_electrons_1960} given by,
\begin{equation}
\frac{1}{\Lambda_{\omega, \mathrm{eff}}} = \frac{1}{\Lambda_{\omega, \mathrm{bulk}}} + \frac{1-p_\omega}{1+p_\omega}\frac{1}{d}
\end{equation}
where $d$ is the thickness of the thin film and $\Lambda_{\omega, \mathrm{bulk}}$ is the intrinsic phonon mean free path in the bulk material. Although the Matthiessen's rule has been used in the past for computational ~\cite{dechaumphai_thermal_2012} and experimental~\cite{hochbaum_enhanced_2008} investigations of phonon boundary scattering, the mathematical rigor of such an expression for effective mean free path is unclear. On the other hand, the effective mean free path of phonons in thin films can also be determined rigorously from the Fuchs-Sondheimer factor ($F\left(\Lambda_\omega/d\right)$), since by definition, $F\left(\Lambda_\omega/d\right) = k_{\omega, \mathrm{eff}}/k_{\omega, \mathrm{bulk}} = \Lambda_{\omega, \mathrm{eff}}/\Lambda_{\omega, \mathrm{bulk}}$. Figure~\ref{fig:Steady} (c) shows the comparison of the normalized effective phonon mean free paths obtained from the Fuchs-Sondheimer factor and Matthiessen's rule for different film thicknesses. Matthiessen's rule underpredicts phonon MFPs comparable to the thickness of the film. Even for phonons with intrinsic mean free path an order of magnitude smaller than the film thickness, Matthiessen's rule predicts a shorter effective phonon mean free path compared to the predictions of the Fuchs-Sondheimer factor from the rigorous solution of the BTE, which is consistent with the findings of another work based on Monte Carlo sampling~\cite{mcgaughey_nanostructure_2012}. This result highlights the importance of using the rigorous BTE solution to estimate the extent of diffuse phonon boundary scattering even in simple nanostructures.
\subsection{Transient Transport in Thin Films}
We now examine transient thermal conduction along thin films observed in the TG experiment. To perform this calculation, we solve the integral equation (equation~\ref{fredholm_2nd_maintxt}) semi-analytically using the same isotropic phonon properties used in steady state transport calculations. The source term in the BTE (equation~\ref{BTE_FT}) is assumed to follow a thermal distribution given by $Q_\omega = C_\omega \Delta T_0$, where $C_\omega$ is the volumetric specific heat of the phonon mode.
\subsubsection{Difference between Thermalizing and Non-thermalizing Boundary Scattering}
\begin{figure}[!ht]
\begin{center}
\includegraphics*[scale=0.5, trim={88mm 115mm 100mm 125mm}, clip]{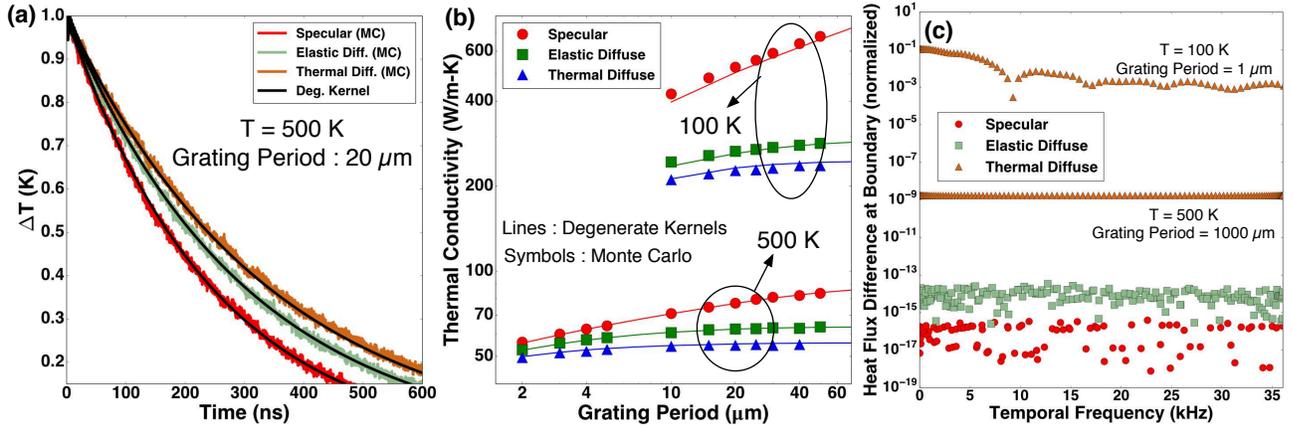}
\end{center}
\caption{(a) Comparison between time traces from the Monte Carlo (colored noisy lines) and the degenerate kernels solutions (black lines) of the BTE for a grating period of $20\ \mu m$ at 500 K. (b) Comparison of the thermal conductivity predictions from the Monte Carlo (symbols) and the degenerate kernels solutions (black lines) of the BTE for different temperatures and grating periods. For both (a) and (b), the Monte Carlo solutions and the BTE solutions from this work are in very good agreement. (c) Plot showing the heat flux conservation at the film boundaries. Specular and non-thermalizing diffuse boundary conditions conserve heat flux to numerical precision while thermalizing diffuse boundary condition violates heat flux conservation at the film wall under quasiballistic (T = 100 K, grating period = $1\ \mu$m) and diffusive (T = 500 K, grating period = $1000\ \mu$m) transport regimes.} \label{fig:Validation_Tvst}
\end{figure}

Figure~\ref{fig:Validation_Tvst} (a) shows a comparison of the time traces calculated from the degenerate kernel method and the Monte Carlo method for a grating period of 20 $\mu$m. The transient decays are in good agreement between the degenerate kernel and the Monte Carlo solutions over a wide range of temperatures and different boundary conditions. As expected, the solution for the specular boundary condition results in a faster transient decay than the diffuse boundary conditions since a specularly reflecting wall does not resist the flow of heat in the in-plane direction. However, the transient decay for the non-thermalizing diffuse boundary condition is faster that the thermalizing diffuse boundary condition, indicating that the thermalizing boundary condition offers higher resistance to heat flow than the non-thermalizing diffuse scattering.\\

This observation is also evident from figure~\ref{fig:Validation_Tvst} (b) which shows the thermal conductivities obtained by fitting the time traces to an exponential decay for different temperatures, different grating periods and different boundary conditions. The observed thermal conductivity of the thin film decreases with decreasing grating period due to the breakdown of the Fourier's law of heat conduction and the onset of quasiballistic thermal transport~\cite{hua_transport_2014} when the grating period is comparable to phonon MFPs. Consistent with the findings from the time traces, the thermal conductivity of the thin film with specular walls is higher than that of the thin film with diffuse walls. Moreover, even for very long grating periods compared to phonon MFPs, where the thermal transport is diffusive and obeys Fourier's law, the thermal conductivity of thin film with non-thermalizing diffuse walls is higher than that of the thin films with thermalizing diffuse walls. This observation is in stark contrast with the steady state condition, where there was no difference in thermal conductivity between thermalizing and non-thermalizing boundary conditions. 
\subsubsection{Validity of the Thermalizing and Non-thermalizing Boundary Conditions}
At this point, it is important to investigate the validity of the thermalizing and non-thermalizing boundary condition for the thin film walls. The non-thermalizing boundary scattering condition can be naturally derived from the conservation of heat flux at the boundary~\cite{zeng_phonon_2000}. However, the thermalizing boundary condition is not derived from the heat flux conservation at the boundary. Therefore, in the absence of any external scattering mechanisms, phonons cannot reach the local thermal equilibrium and simultaneously conserve heat flux at the boundary in general, due to the following reason.\\

Consider a boundary at $z=0$ separating a solid at $z>0$ from vacuum in $z<0$. The incoming phonon distribution at $z=0$ is $g^{-}_\omega\left(0, \mu, \phi\right)$, which is a general phonon distribution, not necessarily at the local thermal equilibrium. According to the formulation of the thermalizing diffuse boundary condition, the outgoing phonon distribution, in the case of fully diffuse boundary scattering, is given by $g_0\left(\Delta T\left(z=0\right)\right)\approx\frac{C_\omega}{4\pi}\Delta T\left(z=0\right)$, where $C_\omega$ is the heat capacity of the phonon mode and $\Delta T\left(z=0\right)$ is the local equilibrium temperature at the boundary $z=0$. Since the boundary separates a solid from vacuum, all of the heat flux incident on the boundary has to be reflected back into the solid. This constraint on the incident and reflected heat flux at the thermalizing diffuse boundary leads to the following relation for $\Delta T\left(z=0\right)$.
\begin{equation}
\Delta T\left(z=0\right) = 4\frac{\sum_{p}\int_{\omega = 0}^{\omega_{\mathrm{max}}}\int_{\mu=-1}^0\int_{\phi=0}^{2\pi}g^{-}_\omega\left(0, \mu, \phi\right)v_g\mu\mathrm{d}\mu\mathrm{d}\phi\mathrm{d}\omega}{\sum_{p}\int_{\omega = 0}^{\omega_{\mathrm{max}}}C_\omega v_g\mathrm{d}\omega} \label{HF_DeltaT}
\end{equation}
Additionally, energy conservation (equation~\ref{e_conserve}) has to be satisfied at all locations including the boundaries in the absence of any other source or sink of phonons. This requirement further adds constraints on $\Delta T\left(z=0\right)$ through the relation,
\begin{equation}
\Delta T\left(z=0\right) = 2\frac{\sum_{p}\int_{\omega = 0}^{\omega_{\mathrm{max}}}\int_{\mu=-1}^0\int_{\phi=0}^{2\pi}\frac{g^{-}_\omega\left(0, \mu, \phi\right)}{\tau_\omega}\mathrm{d}\mu\mathrm{d}\phi\mathrm{d}\omega}{\sum_{p}\int_{\omega = 0}^{\omega_{\mathrm{max}}}\frac{C_\omega}{\tau_\omega}\mathrm{d}\omega} \label{ECons_DeltaT}
\end{equation}
For the assumptions made in the Fuchs-Sondheimer theory under steady state transport conditions, the integrals of the incoming and the outgoing distribution functions (equation~\ref{Steady_gen_soln_2}) over the azimuthal angle $\phi$ are $0$. Therefore, there is no heat flux towards or away from the boundary and the constraints on $\Delta T\left(z=0\right)$ (given by equations~\ref{HF_DeltaT} and~\ref{ECons_DeltaT}) are trivially satisfied. However, in general, these two expressions for $\Delta T\left(z=0\right)$ are not equal, indicating phonons cannot thermalize at the boundaries in the absence of any external source or sink of phonons.\\

 Figure~\ref{fig:Validation_Tvst} (c) shows the difference between the incoming and outgoing total heat flux at the thin film wall ($z=0$) as a function of the temporal frequency $\eta$. The specular and non-thermalizing diffuse boundary conditions satisfy heat flux conservation to numerical precision. However, there is a significant difference between the incoming and the outgoing heat flux for the thermalizing diffuse boundary condition under quasiballistic (T = 100 K, grating period = $1\ \mu$m) and diffusive (T = 500 K, grating period = $1000\ \mu$m) transport regimes. Nevertheless, it is still possible for inelastic (but not thermalizing) diffuse boundary scattering to take place as long as the following conditions for heat flux are met at the thin film boundaries:
\begin{align*}
\begin{split}
\sum_p\int_{\omega = 0}^{\omega_\mathrm{max}}g^+_\omega\left(z=0\right)v_g\mathrm{d}\omega &= -\frac{1}{\pi}\sum_p\int_{\omega = 0}^{\omega_\mathrm{max}}\int_{\mu=-1}^0\int_{\phi=0}^{2\pi}g^-_\omega\left(z=0, \mu, \phi\right)v_g\mu\mathrm{d}\mu\mathrm{d}\phi\mathrm{d}\omega\\
\sum_p\int_{\omega = 0}^{\omega_\mathrm{max}}g^-_\omega\left(z=d\right)v_g\mathrm{d}\omega &= \frac{1}{\pi}\sum_p\int_{\omega = 0}^{\omega_\mathrm{max}}\int_{\mu=0}^1\int_{\phi=0}^{2\pi}g^+_\omega\left(z=d, \mu, \phi\right)v_g\mu\mathrm{d}\mu\mathrm{d}\phi\mathrm{d}\omega\\
\end{split}
\end{align*}
\subsubsection{Comparison with Fuchs-Sondheimer Theory at Different Grating Periods}
\begin{figure}[!ht]
\begin{center}
\includegraphics*[scale=0.575, trim={95mm 100mm 90mm 110mm}, clip]{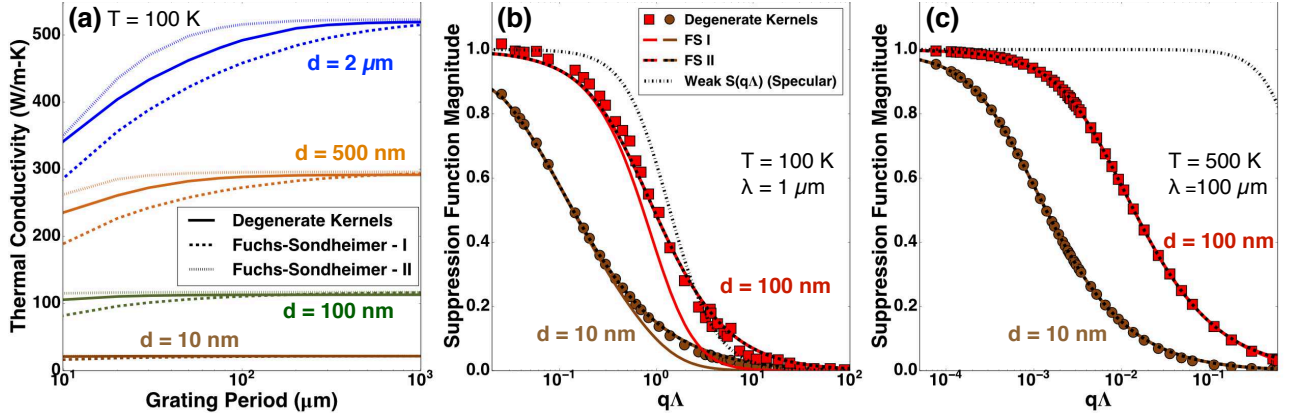}
\end{center}
\caption{Comparison of the thermal conductivity (a) and the suppression functions ((b) and (c)) calculated from the models FS I, FS II and by solving the BTE for non-thermalizing diffuse boundary conditions at different temperatures, grating periods ($\lambda$) and film thicknesses. In figures (b) and (c), the symbols correspond to the degenerate kernel solution, the solid lines correspond to FS I model and the dashed solid lines correspond to FS II model. For very thin films and long grating periods, the models FS I and FS II are in good agreement with the BTE predictions. For thicker films and shorter grating periods, FS I underpredicts and FS II overpredicts the thermal conductivity at short grating periods (a) and the contribution of phonons with long MFP ((b) and (c)) compared to the complete BTE solution.} \label{fig:SuppressionMag_Diffuse}
\end{figure}

We now examine if the Fuchs-Sondheimer theory can be used to explain transient heat conduction in the TG experiment along thin films. If the suppression in thermal conductivity of thin films due to phonon boundary scattering and quasiballistic effects in the TG experiment are assumed to be independent, Fuchs-Sondheimer theory can be employed to describe quasiballistic transport in the TG experiment using the following expression:
\begin{equation}
k\left( q , d\right) = \sum_p\int_0^{\omega_{\mathrm{max}}}F\left(p_\omega, \frac{\Lambda_\omega}{d}\right)S\left( q \Lambda_\omega\right)\left[\frac{1}{3}C_\omega v_g\Lambda_\omega\right]\mathrm{d}\omega \label{FS_Transient1}
\end{equation}
where $ F\left(p_\omega, \frac{\Lambda_\omega}{d}\right)$ is the Fuchs-Sondheimer suppression function from the steady state transport condition and $S\left( q \Lambda_\omega\right)$ is the quasiballistic suppression function~\cite{hua_transport_2014} for a grating period $ q $. Recent works~\cite{ johnson_direct_2013} have used a similar expression for the thermal conductivity suppression of the form:
\begin{equation}
k\left( q , d\right) = \sum_p\int_0^{\omega_{\mathrm{max}}}F\left(p_\omega, \frac{\Lambda_\omega}{d}\right)S\left( q \Lambda_\omega F\left(p_\omega, \frac{\Lambda_\omega}{d}\right)\right)\left[\frac{1}{3}C_\omega v_g\Lambda_\omega\right]\mathrm{d}\omega \label{FS_Transient2}
\end{equation}
Henceforth, equation~\ref{FS_Transient1} is referred to as FS I and equation~\ref{FS_Transient2} is referred to as FS II. Figure~\ref{fig:SuppressionMag_Diffuse} (a) shows the comparison of thermal conductivity obtained by fitting the BTE solution for temperature decay, and thermal conductivities from FS I and FS II models for fully diffuse boundary scattering. We only consider non-thermalizing diffuse scattering as we have shown that thermalizing diffuse scattering is unphysical for the problem considered here. At very long grating periods, when the transport is primarily diffusive, the thermal conductivity predictions from FS I and FS II match well with the BTE solution from this work, as expected. However, at the shorter grating periods comparable to phonon MFPs, where the transport is in the quasiballistic regime, FS I underpredicts the thin film thermal conductivity while FS II overpredicts it.\\

This observation is also evident from the magnitude of the suppression function plotted at $\eta = 0$ for fully diffuse boundary conditions shown in figures~\ref{fig:SuppressionMag_Diffuse} (b) and (c). The suppression function for the thin film geometry is defined as
\begin{equation}
S\left( q \Lambda_\omega, \Lambda_\omega/d, \eta\tau_\omega, p_\omega\right) = \frac{\kappa_{\omega, \mathrm{BTE}}}{\kappa_{\omega, \mathrm{Fourier}}}
\end{equation}
where, $\kappa_\omega = j_{x, \omega}/\Delta \bar{T}$ is the conductance per phonon mode and $j_{x, \omega}$ is the thickness-averaged in-plane heat flux defined in equation~\ref{Heat_flux}. In figures~\ref{fig:SuppressionMag_Diffuse} (b) and~\ref{fig:SuppressionMag_Diffuse} (c), the magnitude of the suppression function at $\eta=0$ is plotted against phonon MFP non-dimensionalized with respect to the grating period $q$. The suppression functions from the complete BTE solution and the models FS I and FS II are identical at high temperatures and long grating periods, when the transport is primarily diffusive, governed by the Fourier's law of heat conduction. However, for low temperatures and short grating periods, FS I underpredicts the heat flux and FS II overpredicts the heat flux carried by phonons with very long MFPs. Moreover, the difference between the models FS I and FS II, and the BTE solution is smaller for thinner films indicating that enhanced boundary scattering in thinner films delays the onset of quasiballistic heat conduction. These observations emphasize the importance of using the complete BTE solution to accurately investigate boundary scattering when grating periods are comparable to phonon MFPs.
\section{Conclusion}
We have studied the effect of thermalizing and non-thermalizing boundary scattering of phonons in steady state and transient heat conduction along thin films by solving the BTE using analytical and computationally efficient semi-analytical techniques. From our analysis, we reach the following conclusions. First, under steady state transport conditions, we find that the thermal transport is governed by the Fuchs-Sondheimer theory and is insensitive to whether the boundaries are thermalizing or not. In contrast, under transient conditions, the decay rates are significantly different for thermalizing and non-thermalizing walls and the Fuchs-Sondheimer theory is only applicable in the heat diffusion regime. We also show that, for transient transport, the thermalizing wall boundary condition is unphysical due to violation of heat flux conservation. Our results provide insights into the boundary scattering process of thermal phonons over a wide range of heating length scales that are useful for interpreting thermal measurements on nanostructures.
\section{Acknowledgments}
Navaneetha K. Ravichandran would like to thank the Resnick Sustainability Institute at Caltech and the Dow Chemical Company for fellowship support. Austin J. Minnich was supported by the National Science Foundation under Grant No. CBET CAREER 1254213.
\bibliographystyle{apsrev4-1}
%

\end{document}


\title{The Role of Thermalizing and Non-thermalizing Walls in Phonons Heat Conduction along Thin Films\\
\underline{Supplementary Information}}
\author{Navaneetha K. Ravichandran}
\author{Austin J. Minnich}
\email{aminnich@caltech.edu}
\affiliation{Division of Engineering and Applied Science,\\ California Institute of Technology, Pasadena, California 91125, USA}
\maketitle
In this report we provide the details of the solution method of the BTE (equation 1 in the main article) for the thin film geometry. Specifically, this supplementary material contains the following information:
\begin{enumerate}
\item In section~\ref{app:SS_Soln}, we describe the complete derivation of the distribution function $\bar{g}_\omega$ from the BTE for steady state transport.
\item In section~\ref{SI:Transient}, we provide a detailed description of all the steps necessary to semi-analytically solve the BTE in transient transport. Section~\ref{appendix_BC} describes the discretization of the boundary conditions, section~\ref{appendix_fredholm} describes the formulation of the integral equation for the temperature distribution $\Delta\bar{T}$ in the frequency domain, and section~\ref{appendix_degenerate_kernels} describes the derivation of different Fourier coefficients to solve the integral equation for $\Delta\bar{T}$ using the method of degenerate kernels.
\item Finally, we describe the Monte Carlo solution technique used to validate the semi-analytical BTE solution in section~\ref{MC_BTE}.
\end{enumerate}
\section{BTE Solution for Steady State Heat Conduction}\label{app:SS_Soln}
In this section, we provide the details of the steps between equations 4 and 7 in the main article.
Under the assumptions of steady state heat conduction consistent with the Fuchs-Sondheimer theory, the BTE becomes,
\begin{equation}
v_g\mu\frac{\partial g_\omega}{\partial z} + v_g\sqrt{1-\mu^2}\cos\phi\frac{\partial g_\omega}{\partial x} = -\frac{g_\omega - g_\omega^0}{\tau_\omega}
\end{equation}
Let $\bar{g}_\omega = g_\omega - g_\omega^0$ represent the deviation from the equilibrium distribution. We further assume that the in-plane gradient of $\bar{g}_\omega$ is small and can be neglected. In this case, the BTE can be simplified as,
\begin{equation}
\frac{\partial \bar{g}_\omega}{\partial z} + \frac{\bar{g}_\omega}{\mu\Lambda_\omega} = -\frac{\cos\phi\sqrt{1-\mu^2}}{\mu}\frac{\partial g^0_\omega}{\partial x}
\end{equation}
which represents a one-dimensional ordinary differential equation whose general solution is given by,
\begin{align}
\label{app:Steady_gen_soln_1}
\begin{split}
\mathrm{For}\ \mu\in\left(0, 1\right]&,\\
\bar{g}_\omega^+\left(z, \mu, \phi\right) &= \bar{g}_\omega^+\left(0, \mu, \phi\right)\exp\left(-\frac{z}{\mu\Lambda_\omega}\right) - \Lambda_\omega\cos\phi\sqrt{1-\mu^2}\frac{\partial g^0_\omega}{\partial x}\left(1-\exp\left(-\frac{z}{\mu\Lambda_\omega}\right)\right)\\
\mathrm{For}\ \mu\in\left[-1, 0\right)&,\\
\bar{g}_\omega^-\left(z, \mu, \phi\right) &= \bar{g}_\omega^-\left(d, \mu, \phi\right)\exp\left(\frac{d-z}{\mu\Lambda_\omega}\right) - \Lambda_\omega\cos\phi\sqrt{1-\mu^2}\frac{\partial g^0_\omega}{\partial x}\left(1-\exp\left(\frac{d-z}{\mu\Lambda_\omega}\right)\right)
\end{split}
\end{align}
The boundary conditions (equation 2 in the main article) for $\bar{g}_\omega$ now become,
\begin{align}
\label{app:Steady_gen_BCs_1}
\begin{split}
\mathrm{For}\ \mu\in\left(0, 1\right]&,\\
\bar{g}_\omega ^+\left(0, \mu, \phi\right) &= p_\omega \bar{g}_\omega ^-\left(0, -\mu, \phi\right) - \frac{\left(1- p_\omega \right)\left(1- \sigma_\omega \right)}{\pi}\int_0^{2\pi}\int_{-1}^0\bar{g}_\omega^-\left(0, \mu', \phi\right)\mu'\mathrm{d}\mu'\mathrm{d}\phi\\
\mathrm{For}\ \mu\in\left[-1, 0\right)&,\\
\bar{g}_\omega ^-\left(d, \mu, \phi\right) &= p_\omega \bar{g}_\omega ^+\left(d, -\mu, \phi\right) + \frac{\left(1- p_\omega \right)\left(1- \sigma_\omega \right)}{\pi}\int_0^{2\pi}\int_0^1\bar{g}_\omega^+\left(d, \mu', \phi\right)\mu'\mathrm{d}\mu'\mathrm{d}\phi
\end{split}
\end{align}
Since $g^0_\omega\left(T\right)\left|\right._{z=0}$ and $g^0_\omega\left(T\right)\left|\right._{z=d}$ are independent of the angular variables $\mu$ and $\phi$, the general solution of the BTE (equation~\ref{app:Steady_gen_BCs_1}) can be substituted into the boundary conditions to get,
\begin{align}
\label{app:Steady_gen_BCs_2}
\begin{split}
\mathrm{For}\ \mu\in\left(0, 1\right]&,\\
\bar{g}_\omega ^+\left(0, \mu, \phi\right) &= p_\omega \bar{g}_\omega^-\left(d, -\mu, \phi\right)\exp\left(-\frac{d}{\mu\Lambda_\omega}\right) - p_\omega \Lambda_\omega\cos\phi\sqrt{1-\mu^2}\frac{\partial g^0_\omega}{\partial x}\left(1-\exp\left(-\frac{d}{\mu\Lambda_\omega}\right)\right)\\
&\ \ \ \ \ - \frac{(1- p_\omega )(1- \sigma_\omega )}{\pi}\int_0^{2\pi}\int_{-1}^0\bar{g}_\omega^-\left(d, \mu', \phi\right)\exp\left(\frac{d}{\mu'\Lambda_\omega}\right)\mu'\mathrm{d}\mu'\mathrm{d}\phi \\
&\ \ \ \ \ + \frac{(1- p_\omega )(1- \sigma_\omega )}{\pi}\int_0^{2\pi}\int_{-1}^0\Lambda_\omega\cos\phi\sqrt{1-\mu'^2}\frac{\partial g^0_\omega}{\partial x}\left(1-\exp\left(\frac{d}{\mu'\Lambda_\omega}\right)\right)\mu'\mathrm{d}\mu'\mathrm{d}\phi\\
&= p_\omega \bar{g}_\omega^-\left(d, -\mu, \phi\right)\exp\left(-\frac{d}{\mu\Lambda_\omega}\right) - p_\omega \Lambda_\omega\cos\phi\sqrt{1-\mu^2}\frac{\partial g^0_\omega}{\partial x}\left(1-\exp\left(-\frac{d}{\mu\Lambda_\omega}\right)\right)\\
&\ \ \ \ \ - \frac{(1- p_\omega )(1- \sigma_\omega )}{\pi}\int_0^{2\pi}\int_{-1}^0\bar{g}_\omega^-\left(d, \mu', \phi\right)\exp\left(\frac{d}{\mu'\Lambda_\omega}\right)\mu'\mathrm{d}\mu'\mathrm{d}\phi \\
\mathrm{For}\ \mu\in\left[-1, 0\right)&,\\
\bar{g}_\omega ^-\left(d, \mu, \phi\right) &= p_\omega \bar{g}_\omega^+\left(0, -\mu, \phi\right)\exp\left(\frac{d}{\mu\Lambda_\omega}\right) - p_\omega \Lambda_\omega\cos\phi\sqrt{1-\mu^2}\frac{\partial g^0_\omega}{\partial x}\left(1-\exp\left(\frac{d}{\mu\Lambda_\omega}\right)\right)\\
&\ \ \ \ \ + \frac{(1- p_\omega )(1- \sigma_\omega )}{\pi}\int_0^{2\pi}\int_0^1\bar{g}_\omega^+\left(0, \mu', \phi\right)\exp\left(-\frac{d}{\mu'\Lambda_\omega}\right)\mu'\mathrm{d}\mu'\mathrm{d}\phi \\
&\ \ \ \ \ - \frac{(1- p_\omega )(1- \sigma_\omega )}{\pi}\int_0^{2\pi}\int_0^1\Lambda_\omega\cos\phi\sqrt{1-\mu'^2}\frac{\partial g^0_\omega}{\partial x}\left(1-\exp\left(-\frac{d}{\mu'\Lambda_\omega}\right)\right)\mu'\mathrm{d}\mu'\mathrm{d}\phi\\
&= p_\omega \bar{g}_\omega^+\left(0, -\mu, \phi\right)\exp\left(\frac{d}{\mu\Lambda_\omega}\right) - p_\omega \Lambda_\omega\cos\phi\sqrt{1-\mu^2}\frac{\partial g^0_\omega}{\partial x}\left(1-\exp\left(\frac{d}{\mu\Lambda_\omega}\right)\right)\\
&\ \ \ \ \ + \frac{(1- p_\omega )(1- \sigma_\omega )}{\pi}\int_0^{2\pi}\int_0^1\bar{g}_\omega^+\left(0, \mu', \phi\right)\exp\left(-\frac{d}{\mu'\Lambda_\omega}\right)\mu'\mathrm{d}\mu'\mathrm{d}\phi \\
\end{split}
\end{align}
since $\int_0^{2\pi}\cos\phi\mathrm{d}\phi = 0$. For simplicity and convenience, we change the limits of the variables $\mu$ and $\mu'$ from $\left[-1, 0\right)$ to $\left(0, 1\right]$ wherever necessary to get,
\begin{align}
\label{app:Steady_gen_BCs_3}
\begin{split}
\bar{g}_\omega ^+\left(0, \mu, \phi\right) &= p_\omega \bar{g}_\omega^-\left(d, -\mu, \phi\right)\exp\left(-\frac{d}{\mu\Lambda_\omega}\right) - p_\omega \Lambda_\omega\cos\phi\sqrt{1-\mu^2}\frac{\partial g^0_\omega}{\partial x}\left(1-\exp\left(-\frac{d}{\mu\Lambda_\omega}\right)\right)\\
&\ \ \ \ \ + \left(1- p_\omega \right)\left(1- \sigma_\omega \right)A^+_\omega \\
\bar{g}_\omega ^-\left(d, -\mu, \phi\right) &= p_\omega \bar{g}_\omega^+\left(0, \mu, \phi\right)\exp\left(-\frac{d}{\mu\Lambda_\omega}\right) - p_\omega \Lambda_\omega\cos\phi\sqrt{1-\mu^2}\frac{\partial g^0_\omega}{\partial x}\left(1-\exp\left(-\frac{d}{\mu\Lambda_\omega}\right)\right)\\
&\ \ \ \ \ + \left(1- p_\omega \right)\left(1- \sigma_\omega \right)A^-_\omega
\end{split}
\end{align}
where, $A^+_\omega$ and $A^-_\omega$ are constants, independent of the angular variables $\mu$ and $\phi$, given by,
\begin{align*}
\begin{split}
A^+_\omega &= \frac{1}{\pi}\int_0^{2\pi}\int_0^1\bar{g}_\omega^-\left(d, -\mu', \phi\right)\exp\left(-\frac{d}{\mu'\Lambda_\omega}\right)\mu'\mathrm{d}\mu'\mathrm{d}\phi\\
A^-_\omega &= \frac{1}{\pi}\int_0^{2\pi}\int_0^1\bar{g}_\omega^+\left(0, \mu', \phi\right)\exp\left(-\frac{d}{\mu'\Lambda_\omega}\right)\mu'\mathrm{d}\mu'\mathrm{d}\phi
\end{split}
\end{align*}
Solving these boundary conditions (equation~\ref{app:Steady_gen_BCs_3}), we get,
\begin{align}
\label{app:Steady_gen_BCs_4}
\begin{split}
\bar{g}_\omega ^+\left(0, \mu, \phi\right) &= -\ p_\omega\Lambda_\omega\cos\phi\sqrt{1-\mu^2}\frac{\partial g^0_\omega}{\partial x}\frac{\left(1-\exp\left(-\frac{d}{\mu\Lambda_\omega}\right)\right)}{1- p_\omega \exp\left(-\frac{d}{\mu\Lambda_\omega}\right)} \\
&\ \ \ \ \ + \frac{\left(1- p_\omega \right)\left(1- \sigma_\omega \right)\left[A^+_\omega + p_\omega \exp\left(-\frac{d}{\mu\Lambda_\omega}\right)A^-_\omega\right]}{1 - p^2_\omega \exp\left(-\frac{2d}{\mu\Lambda_\omega}\right)}\\
\bar{g}_\omega ^-\left(d, -\mu, \phi\right) &= -\ p_\omega\Lambda_\omega\cos\phi\sqrt{1-\mu^2}\frac{\partial g^0_\omega}{\partial x}\frac{\left(1-\exp\left(-\frac{d}{\mu\Lambda_\omega}\right)\right)}{1- p_\omega \exp\left(-\frac{d}{\mu\Lambda_\omega}\right)} \\
&\ \ \ \ \ + \frac{\left(1- p_\omega \right)\left(1- \sigma_\omega \right)\left[A^-_\omega + p_\omega \exp\left(-\frac{d}{\mu\Lambda_\omega}\right)A^+_\omega\right]}{1 - p^2_\omega \exp\left(-\frac{2d}{\mu\Lambda_\omega}\right)}\\
\end{split}
\end{align}
Therefore, the general solution can now be written as,
\begin{align}
\label{app:Steady_gen_soln_2}
\begin{split}
\bar{g}_\omega^+\left(z, \mu, \phi\right) &= -\ \Lambda_\omega\cos\phi\sqrt{1-\mu^2}\frac{\partial g_\omega^0}{\partial x}\left[\frac{p_\omega\left(1-\exp\left(-\frac{d}{\mu\Lambda_\omega}\right)\right)\exp\left(-\frac{z}{\mu\Lambda_\omega}\right)}{1- p_\omega \exp\left(-\frac{d}{\mu\Lambda_\omega}\right)}
+ \left(1-\exp\left(-\frac{z}{\mu\Lambda_\omega}\right)\right)\right] \\
&\ \ \ \ \ + \frac{\left(1- p_\omega \right)\left(1- \sigma_\omega \right)\left[A^+_\omega + p_\omega \exp\left(-\frac{d}{\mu\Lambda_\omega}\right)A^-_\omega\right]}{1 - p^2_\omega \exp\left(-\frac{2d}{\mu\Lambda_\omega}\right)}\exp\left(-\frac{z}{\mu\Lambda_\omega}\right)\\
&= -\ \Lambda_\omega\cos\phi\sqrt{1-\mu^2}\frac{\partial g_\omega^0}{\partial x}\Bigg[1-\frac{\left(1-p_\omega\right)\exp\left(-\frac{z}{\mu\Lambda_\omega}\right)}{1- p_\omega\exp\left(-\frac{d}{\mu\Lambda_\omega}\right)}\Bigg] \\
&\ \ \ \ \ + \frac{\left(1- p_\omega \right)\left(1- \sigma_\omega \right)\left[A^+_\omega + p_\omega \exp\left(-\frac{d}{\mu\Lambda_\omega}\right)A^-_\omega\right]}{1 - p^2_\omega \exp\left(-\frac{2d}{\mu\Lambda_\omega}\right)}\exp\left(-\frac{z}{\mu\Lambda_\omega}\right)\\
\bar{g}_\omega^-\left(z, -\mu, \phi\right) &= -\Lambda_\omega\cos\phi\sqrt{1-\mu^2}\frac{\partial g_\omega^0}{\partial x}\Bigg[\frac{ p_\omega\left(1-\exp\left(-\frac{d}{\mu\Lambda_\omega}\right)\right)\exp\left(-\frac{\left(d-z\right)}{\mu\Lambda_\omega}\right)}{1- p_\omega \exp\left(-\frac{d}{\mu\Lambda_\omega}\right)} + \left(1-\exp\left(-\frac{\left(d-z\right)}{\mu\Lambda_\omega}\right)\right)\Bigg] \\
&\ \ \ \ \ + \frac{\left(1- p_\omega \right)\left(1- \sigma_\omega \right)\left[A^-_\omega + p_\omega \exp\left(-\frac{d}{\mu\Lambda_\omega}\right)A^+_\omega\right]}{1 - p^2_\omega \exp\left(-\frac{2d}{\mu\Lambda_\omega}\right)}\exp\left(-\frac{\left(d-z\right)}{\mu\Lambda_\omega}\right)\\
&= -\Lambda_\omega\cos\phi\sqrt{1-\mu^2}\frac{\partial g_\omega^0}{\partial x}\Bigg[1-\frac{\left(1- p_\omega \right)\exp\left(-\frac{\left(d-z\right)}{\mu\Lambda_\omega}\right)}{1- p_\omega \exp\left(-\frac{d}{\mu\Lambda_\omega}\right)}\Bigg] \\
&\ \ \ \ \ + \frac{\left(1- p_\omega \right)\left(1- \sigma_\omega \right)\left[A^-_\omega + p_\omega \exp\left(-\frac{d}{\mu\Lambda_\omega}\right)A^+_\omega\right]}{1 - p^2_\omega \exp\left(-\frac{d}{\mu\Lambda_\omega}\right)}\exp\left(-\frac{\left(d-z\right)}{\mu\Lambda_\omega}\right)\\
\end{split}
\end{align}
As described in the main article, we substitute these general solutions for $\bar{g}_\omega$ into the expression for heat flux (equation 8 in the main article) and derive the suppression in thermal conductivity due to phonon boundary scattering. Since $A^{+}_\omega$ and $A^{-}_\omega$ in equation~\ref{app:Steady_gen_soln_2} are independent of the angular coordinates $\mu$ and $\phi$, the only terms containing the thermalization parameter $\sigma_\omega$ in $\bar{g}_\omega$ integrate out to $0$ while evaluating the thermal conductivity of the thin film. Therefore, steady state thermal conductivity measurements in thin films cannot be used to distinguish between thermalizing and non-thermalizing phonon boundary scattering. 
\section{BTE Solution for Transient Heat Conduction} \label{SI:Transient}
In this section, we discuss different parts of the BTE solution methodology for the transient transport condition. Under this section, we describe the discretization of the boundary conditions in section~\ref{appendix_BC},  the formulation of the integral equation for the temperature distribution $\Delta\bar{T}$ in the frequency domain in section~\ref{appendix_fredholm} and the derivation of different Fourier coefficients to solve the integral equation for $\Delta\bar{T}$ using the method of degenerate kernels in section~\ref{appendix_degenerate_kernels}.
\subsection{Numerical Discretization of the Boundary Conditions} \label{appendix_BC}
The general boundary conditions at the thin film walls are given by,
\begin{align}
\label{app:bdry_conditions}
\begin{split}
\mathrm{For}\ \mu\in\left(0, 1\right], & \\
g^+_\omega\left(0, \mu, \phi\right) &= p_\omega g^-_\omega\left(0, -\mu, \phi\right) \\
&\ \ \ \ \ \ \ + \left(1- p_\omega \right)\Bigg( \sigma_\omega g^0_\omega\left(\Delta T\left(z=0\right)\right) - \frac{\left(1- \sigma_\omega \right)}{\pi}\int_0^{2\pi}\int_{-1}^0g^-_\omega\left(0, \mu ', \phi'\right)\mu'\mathrm{d}\mu '\mathrm{d}\phi'\Bigg)\\
&= p_\omega g^-_\omega\left(0, -\mu, \phi\right) \\
&\ \ \ \ \ \ \ + \left(1- p_\omega \right)\Bigg( \sigma_\omega \frac{C_\omega\Delta T\left(z=0\right)}{4\pi} - \frac{\left(1- \sigma_\omega \right)}{\pi}\int_0^{2\pi}\int_{-1}^0g^-_\omega\left(0, \mu ', \phi'\right)\mu'\mathrm{d}\mu '\mathrm{d}\phi'\Bigg)\\
\mathrm{For}\ \mu\in\left[-1, 0\right), & \\
g^-_\omega\left(d, \mu, \phi\right) &= p_\omega g^+_\omega\left(d, -\mu, \phi\right) \\
&\ \ \ \ \ \ \ + \left(1- p_\omega \right)\Bigg( \sigma_\omega g^0_\omega\left(\Delta T\left(z=d\right)\right) + \frac{\left(1- \sigma_\omega \right)}{\pi}\int_0^{2\pi}\int_0^1g^+_\omega\left(d, \mu ', \phi'\right)\mu'\mathrm{d}\mu '\mathrm{d}\phi'\Bigg)\\
&= p_\omega g^+_\omega\left(d, -\mu, \phi\right) \\
&\ \ \ \ \ \ \ + \left(1- p_\omega \right)\Bigg( \sigma_\omega \frac{C_\omega\Delta T\left(z=d\right)}{4\pi} + \frac{\left(1- \sigma_\omega \right)}{\pi}\int_0^{2\pi}\int_0^1g^+_\omega\left(d, \mu ', \phi'\right)\mu'\mathrm{d}\mu '\mathrm{d}\phi'\Bigg)
\end{split}
\end{align}
In the frequency domain, the boundary conditions (equation~\ref{app:bdry_conditions}) can be written as,
\begin{align}
\label{app:bdry_conditions_freq}
\begin{split}
\mathrm{For}\ \mu\in\left(0, 1\right], & \\
G^+_\omega\left(0, \mu, \phi\right) &= p_\omega G^-_\omega\left(0, -\mu, \phi\right) \\
&\ \ \ \ \ \ \ + \left(1- p_\omega \right)\Bigg( \sigma_\omega \frac{C_\omega\Delta \bar{T}\left(z=0\right)}{4\pi} - \frac{\left(1- \sigma_\omega \right)}{\pi}\int_0^{2\pi}\int_{-1}^0G^-_\omega\left(0, \mu ', \phi'\right)\mu'\mathrm{d}\mu '\mathrm{d}\phi'\Bigg)\\
\mathrm{For}\ \mu\in\left[-1, 0\right), & \\
G^-_\omega\left(d, \mu, \phi\right) &= p_\omega G^+_\omega\left(d, -\mu, \phi\right) \\
&\ \ \ \ \ \ \ + \left(1- p_\omega \right)\Bigg( \sigma_\omega \frac{C_\omega\Delta \bar{T}\left(z=d\right)}{4\pi} + \frac{\left(1- \sigma_\omega \right)}{\pi}\int_0^{2\pi}\int_0^1G^+_\omega\left(d, \mu ', \phi'\right)\mu'\mathrm{d}\mu '\mathrm{d}\phi'\Bigg)
\end{split}
\end{align}
For any given $\mu$ and $\phi$, there are 4 unknown quantities to be determined at the thin film boundaries: $G^+_\omega\left(0, \mu, \phi\right)$, $G^-_\omega\left(0, -\mu, \phi\right)$, $G^+_\omega\left(d, -\mu, \phi\right)$ and $G^-_\omega\left(d, \mu, \phi\right)$, while there are only two equations which are directly evident (equation~\ref{app:bdry_conditions_freq}). However, closed-form relations for these 4 unknown quantities can be obtained in terms of the unknown temperature distribution at the thin film boundaries in the frequency domain ($\Delta \bar{T}\left(z=0\right)$ and $\Delta \bar{T}\left(z=d\right)$) by substituting the general solution of the BTE (equation 11 in the main article) into boundary conditions (equation~\ref{app:bdry_conditions_freq}) to get,
\begin{align}
\label{app:bdry_gen_sub_1}
\begin{split}
\mathrm{For}\ \mu\in\left(0, 1\right],\ &\\ G^+_\omega\left(0, \mu, \phi\right) &= p_\omega G^-_\omega\left(d, -\mu, \phi\right)\exp\left(-\frac{\gamma^{\mathrm{FS}}_{\mu\phi}}{\mu\Lambda_\omega}d\right) \\
&\ \ \ \ \ + \frac{ p_\omega }{4\pi\mu\Lambda_\omega}\int_0^d\left(C_\omega\Delta\bar{T} + \bar{Q}_\omega\tau_\omega\right)\exp\left(-\frac{\gamma^{\mathrm{FS}}_{\mu\phi}}{\mu\Lambda_\omega}z'\right)\mathrm{d}z'\\
&\ \ \ \ \ + \left(1- p_\omega \right)\Bigg[ \sigma_\omega \frac{C_\omega\Delta \bar{T}\left(z=0\right)}{4\pi} \\
&\ \ \ \ \ - \frac{\left(1- \sigma_\omega \right)}{\pi}\int_0^{2\pi}\int_{-1}^0G^-_\omega\left(d, \mu ', \phi'\right)\exp\left(\frac{\gamma^{\mathrm{FS}}_{\mu'\phi'}}{\mu '\Lambda_\omega}d\right)\mu'\mathrm{d}\mu '\mathrm{d}\phi'\\
&\ \ \ \ \ + \frac{\left(1- \sigma_\omega \right)}{4\pi^2\Lambda_\omega}\int_0^{2\pi}\int_{-1}^0\int_0^d\left(C_\omega\Delta\bar{T} + \bar{Q}_\omega\tau_\omega\right)\exp\left(\frac{\gamma^{\mathrm{FS}}_{\mu'\phi'}}{\mu '\Lambda_\omega}z'\right)\mathrm{d}z'\mathrm{d}\mu '\mathrm{d}\phi'\Bigg]\\
\mathrm{For}\ \mu\in\left[-1, 0\right),\ &\\ G^-_\omega\left(d, \mu, \phi\right) &= p_\omega G^+_\omega\left(0, -\mu, \phi\right)\exp\left(\frac{\gamma^{\mathrm{FS}}_{\mu\phi}}{\mu\Lambda_\omega}d\right) \\
&\ \ \ \ \ - \frac{ p_\omega \exp\left(\frac{\gamma^{\mathrm{FS}}_{\mu\phi}}{\mu\Lambda_\omega}d\right)}{4\pi\mu\Lambda_\omega}\int_0^d\left(C_\omega\Delta\bar{T} + \bar{Q}_\omega\tau_\omega\right)\exp\left(-\frac{\gamma^{\mathrm{FS}}_{\mu\phi}}{\mu\Lambda_\omega}z'\right)\mathrm{d}z'\\
&\ \ \ \ \ + \left(1- p_\omega \right)\Bigg[ \sigma_\omega \frac{C_\omega\Delta \bar{T}\left(z=d\right)}{4\pi} \\
&\ \ \ \ \ + \frac{\left(1- \sigma_\omega \right)}{\pi}\int_0^{2\pi}\int_0^1G^+_\omega\left(0, \mu ', \phi'\right)\exp\left(-\frac{\gamma^{\mathrm{FS}}_{\mu'\phi'}}{\mu '\Lambda_\omega}d\right)\mu'\mathrm{d}\mu '\mathrm{d}\phi'\\
&\ \ \ \ \ + \frac{\left(1- \sigma_\omega \right)}{4\pi^2\Lambda_\omega}\int_0^{2\pi}\int_0^1\exp\left(-\frac{\gamma^{\mathrm{FS}}_{\mu'\phi'}}{\mu '\Lambda_\omega}d\right)\int_0^d\left(C_\omega\Delta\bar{T} + \bar{Q}_\omega\tau_\omega\right)\\
&\ \ \ \ \ \ \ \ \ \times\exp\left(\frac{\gamma^{\mathrm{FS}}_{\mu'\phi'}}{\mu '\Lambda_\omega}z'\right)\mathrm{d}z'\mathrm{d}\mu '\mathrm{d}\phi'\Bigg]\\
\end{split}
\end{align}
For convenience, the limits on variables $\mu$ and $\mu '$ are changed from $[-1, 1]$ to $(0, 1]$ in equation~\ref{app:bdry_gen_sub_1} wherever necessary to obtain
\begin{align}
\label{app:bdry_gen_sub_2}
\begin{split}
G^+_\omega\left(0, \mu, \phi\right) &= p_\omega G^-_\omega\left(d, -\mu, \phi\right)\exp\left(-\frac{\gamma^{\mathrm{FS}}_{\mu\phi}}{\mu\Lambda_\omega}d\right) \\
&\ \ \ \ \ + \frac{ p_\omega }{4\pi\mu\Lambda_\omega}\int_0^d\left(C_\omega\Delta\bar{T} + \bar{Q}_\omega\tau_\omega\right)\exp\left(-\frac{\gamma^{\mathrm{FS}}_{\mu\phi}}{\mu\Lambda_\omega}z'\right)\mathrm{d}z'\\
&\ \ \ \ \ + \left(1- p_\omega \right)\Bigg[ \sigma_\omega \frac{C_\omega\Delta \bar{T}\left(z=0\right)}{4\pi} \\
&\ \ \ \ \ + \frac{\left(1- \sigma_\omega \right)}{\pi}\int_0^{2\pi}\int_0^1G^-_\omega\left(d, -\mu ', \phi'\right)\exp\left(-\frac{\gamma^{\mathrm{FS}}_{\mu'\phi'}}{\mu '\Lambda_\omega}d\right)\mu'\mathrm{d}\mu '\mathrm{d}\phi'\\
&\ \ \ \ \ + \frac{\left(1- \sigma_\omega \right)}{4\pi^2\Lambda_\omega}\int_0^{2\pi}\int_0^1\int_0^d\left(C_\omega\Delta\bar{T} + \bar{Q}_\omega\tau_\omega\right)\exp\left(-\frac{\gamma^{\mathrm{FS}}_{\mu'\phi'}}{\mu '\Lambda_\omega}z'\right)\mathrm{d}z'\mathrm{d}\mu '\mathrm{d}\phi'\Bigg]\\
G^-_\omega\left(d, -\mu, \phi\right) &= p_\omega G^+_\omega\left(0, \mu, \phi\right)\exp\left(-\frac{\gamma^{\mathrm{FS}}_{\mu\phi}}{\mu\Lambda_\omega}d\right) \\
&\ \ \ \ \ + \frac{ p_\omega }{4\pi\mu\Lambda_\omega}\int_0^d\left(C_\omega\Delta\bar{T} + \bar{Q}_\omega\tau_\omega\right)\exp\left(-\frac{\gamma^{\mathrm{FS}}_{\mu\phi}}{\mu\Lambda_\omega}\left(d-z'\right)\right)\mathrm{d}z'\\
&\ \ \ \ \ + \left(1- p_\omega \right)\Bigg[ \sigma_\omega \frac{C_\omega\Delta \bar{T}\left(z=d\right)}{4\pi} \\
&\ \ \ \ \ + \frac{\left(1- \sigma_\omega \right)}{\pi}\int_0^{2\pi}\int_0^1G^+_\omega\left(0, \mu ', \phi'\right)\exp\left(-\frac{\gamma^{\mathrm{FS}}_{\mu'\phi'}}{\mu '\Lambda_\omega}d\right)\mu'\mathrm{d}\mu '\mathrm{d}\phi'\\
&\ \ \ \ \ + \frac{\left(1- \sigma_\omega \right)}{4\pi^2\Lambda_\omega}\int_0^{2\pi}\int_0^1\int_0^d\left(C_\omega\Delta\bar{T} + \bar{Q}_\omega\tau_\omega\right)\exp\left(-\frac{\gamma^{\mathrm{FS}}_{\mu'\phi'}}{\mu '\Lambda_\omega}\left(d-z'\right)\right)\mathrm{d}z'\mathrm{d}\mu '\mathrm{d}\phi'\Bigg]\\
\end{split}
\end{align}
Equation~\ref{app:bdry_gen_sub_2} represents a system of integral equations to solve for the 2 unknown quantities $G^+_\omega\left(0, \mu, \phi\right)$ and $G^-_\omega\left(d, -\mu, \phi\right)$ for every $\mu$ and $\phi$. To solve this system of equations, the integrals in $\mu'$ and $\phi'$ variables are first discretized using Gauss quadrature,
\begin{equation}
\int_0^{2\pi}\int_0^1 f\left(\mu', \phi'\right)\mathrm{d}\mu'\mathrm{d}\phi' = \sum_{ij}f\left(\mu _i, \phi _j\right)w_{\mu_i}w_{\phi_j} \label{app:gquad}
\end{equation}
where $\mu_i$ and $\phi_j$ are the quadrature points and $w_{\mu_i}$ and $w_{\phi_j}$ are the corresponding weights. Therefore, equation~\ref{app:bdry_gen_sub_2} transforms into,
\begin{align}
\label{app:bdry_gen_sub_3}
\begin{split}
G^+_\omega\left(0, \mu_i, \phi_j\right) &= p_\omega G^-_\omega\left(d, -\mu_i, \phi_j\right)\exp\left(-\frac{\gamma^{\mathrm{FS}}_{ij}}{\mu_i\Lambda_\omega}d\right) \\
&\ \ \ \ \ + \frac{ p_\omega }{4\pi\mu_i\Lambda_\omega}\int_0^d\left(C_\omega\Delta\bar{T} + \bar{Q}_\omega\tau_\omega\right)\exp\left(-\frac{\gamma^{\mathrm{FS}}_{ij}}{\mu_i\Lambda_\omega}z'\right)\mathrm{d}z'\\
&\ \ \ \ \ + \left(1- p_\omega \right)\Bigg[ \sigma_\omega \frac{C_\omega\Delta \bar{T}\left(z=0\right)}{4\pi} \\
&\ \ \ \ \ + \frac{\left(1- \sigma_\omega \right)}{\pi}\sum_{i'j'}G^-_\omega\left(d, -\mu'_i, \phi'_j\right)\exp\left(-\frac{\gamma^{\mathrm{FS}}_{i'j'}}{\mu '_i\Lambda_\omega}d\right)\mu'_iw_{\mu'_i}w_{\phi'_j}\\
&\ \ \ \ \ + \frac{\left(1- \sigma_\omega \right)}{4\pi^2\Lambda_\omega}\sum_{i'j'}\int_0^d\left(C_\omega\Delta\bar{T} + \bar{Q}_\omega\tau_\omega\right)\exp\left(-\frac{\gamma^{\mathrm{FS}}_{i'j'}}{\mu '_i\Lambda_\omega}z'\right)\mathrm{d}z'w_{\mu'_i}w_{\phi'_j}\Bigg]\\
G^-_\omega\left(d, -\mu_i, \phi_j\right) &= p_\omega G^+_\omega\left(0, \mu_i, \phi_j\right)\exp\left(-\frac{\gamma^{\mathrm{FS}}_{ij}}{\mu_i\Lambda_\omega}d\right) \\
&\ \ \ \ \ + \frac{ p_\omega }{4\pi\mu_i\Lambda_\omega}\int_0^d\left(C_\omega\Delta\bar{T} + \bar{Q}_\omega\tau_\omega\right)\exp\left(-\frac{\gamma^{\mathrm{FS}}_{ij}}{\mu_i\Lambda_\omega}\left(d-z'\right)\right)\mathrm{d}z'\\
&\ \ \ \ \ + \left(1- p_\omega \right)\Bigg[ \sigma_\omega \frac{C_\omega\Delta \bar{T}\left(z=d\right)}{4\pi} \\
&\ \ \ \ \ + \frac{\left(1- \sigma_\omega \right)}{\pi}\sum_{i'j'}G^+_\omega\left(0, \mu'_i, \phi'_j\right)\exp\left(-\frac{\gamma^{\mathrm{FS}}_{i'j'}}{\mu '_i\Lambda_\omega}d\right)\mu'_iw_{\mu'_i}w_{\phi'_j}\\
&\ \ \ \ \ + \frac{\left(1- \sigma_\omega \right)}{4\pi^2\Lambda_\omega}\sum_{i'j'}\int_0^d\left(C_\omega\Delta\bar{T} + \bar{Q}_\omega\tau_\omega\right)\exp\left(-\frac{\gamma^{\mathrm{FS}}_{i'j'}}{\mu'_i\Lambda_\omega}\left(d-z'\right)\right)\mathrm{d}z'w_{\mu'_i}w_{\phi'_j}\Bigg]\\
\end{split}
\end{align}
To simplify these expressions, we substitute the following into equation~\ref{app:bdry_gen_sub_3}:
\begin{align}
\label{app:I_eqns}
\begin{split}
I^+_{\mu\phi} &= \int_0^d\left(C_\omega\Delta\bar{T} + \bar{Q}_\omega\tau_\omega\right)\exp\left(-\frac{\gamma^{\mathrm{FS}}_{\mu\phi}}{\mu\Lambda_\omega}z'\right)\mathrm{d}z'\\
&= C_\omega\int_0^d\Delta\bar{T}\exp\left(-\frac{\gamma^{\mathrm{FS}}_{\mu\phi}}{\mu\Lambda_\omega}z'\right)\mathrm{d}z' + \bar{Q}_\omega\tau_\omega\Lambda_\omega\frac{\mu}{\gamma^{\mathrm{FS}}_{\mu\phi}}\left(1-\exp\left(-\frac{\gamma^{\mathrm{FS}}_{\mu\phi}}{\mu\Lambda_\omega}d\right)\right)\\
I^-_{\mu\phi} &= \int_0^d\left(C_\omega\Delta\bar{T} + \bar{Q}_\omega\tau_\omega\right)\exp\left(-\frac{\gamma^{\mathrm{FS}}_{\mu\phi}}{\mu\Lambda_\omega}\left(d-z'\right)\right)\mathrm{d}z'\\
&= C_\omega\int_0^d\Delta\bar{T}\exp\left(-\frac{\gamma^{\mathrm{FS}}_{\mu\phi}}{\mu\Lambda_\omega}\left(d-z'\right)\right)\mathrm{d}z' + \bar{Q}_\omega\tau_\omega\Lambda_\omega\frac{\mu}{\gamma^{\mathrm{FS}}_{\mu\phi}}\left(1-\exp\left(-\frac{\gamma^{\mathrm{FS}}_{\mu\phi}}{\mu\Lambda_\omega}d\right)\right)\\
\end{split}
\end{align}
which transform equation~\ref{app:bdry_gen_sub_3} into
\begin{align}
\label{app:bdry_gen_sub_4}
\begin{split}
G^+_\omega\left(0, \mu_i, \phi_j\right) &= p_\omega G^-_\omega\left(d, -\mu_i, \phi_j\right)\exp\left(-\frac{\gamma^{\mathrm{FS}}_{ij}}{\mu_i\Lambda_\omega}d\right) + \frac{ p_\omega }{4\pi\mu_i\Lambda_\omega}I^+_{ij}\\
&\ \ \ \ \ + \left(1- p_\omega \right)\Bigg[ \sigma_\omega \frac{C_\omega\Delta \bar{T}\left(z=0\right)}{4\pi} \\
&\ \ \ \ \ + \frac{\left(1- \sigma_\omega \right)}{\pi}\sum_{i'j'}G^-_\omega\left(d, -\mu'_i, \phi'_j\right)\exp\left(-\frac{\gamma^{\mathrm{FS}}_{i'j'}}{\mu '_i\Lambda_\omega}d\right)\mu'_iw_{\mu'_i}w_{\phi'_j}\\
&\ \ \ \ \ + \frac{\left(1- \sigma_\omega \right)}{4\pi^2\Lambda_\omega}\sum_{i'j'}w_{\mu'_i}w_{\phi'_j}I^+_{i'j'}\Bigg]\\
G^-_\omega\left(d, -\mu_i, \phi_j\right) &= p_\omega G^+_\omega\left(0, \mu_i, \phi_j\right)\exp\left(-\frac{\gamma^{\mathrm{FS}}_{ij}}{\mu_i\Lambda_\omega}d\right) + \frac{ p_\omega }{4\pi\mu_i\Lambda_\omega}I^-_{ij}\\
&\ \ \ \ \ + \left(1- p_\omega \right)\Bigg[ \sigma_\omega \frac{C_\omega\Delta \bar{T}\left(z=d\right)}{4\pi} \\
&\ \ \ \ \ + \frac{\left(1- \sigma_\omega \right)}{\pi}\sum_{i'j'}G^+_\omega\left(0, \mu'_i, \phi'_j\right)\exp\left(-\frac{\gamma^{\mathrm{FS}}_{i'j'}}{\mu '_i\Lambda_\omega}d\right)\mu'_iw_{\mu'_i}w_{\phi'_j}\\
&\ \ \ \ \ + \frac{\left(1- \sigma_\omega \right)}{4\pi^2\Lambda_\omega}\sum_{i'j'}w_{\mu'_i}w_{\phi'_j}I^+_{i'j'}\Bigg]\\
\end{split}
\end{align}
These discretized boundary conditions (equation~\ref{app:bdry_gen_sub_4}) can be written in a concise matrix form as
\begin{equation}
[A]\mathbf{G}_{\mathrm{BC}} = \bar{\tilde{\mathbf{c}}} \label{app:bdry_gen_sub_5}
\end{equation}
with the solution of the form
\begin{equation}
\mathbf{G}_{\mathrm{BC}} = [A]^{-1}\bar{\tilde{\mathbf{c}}} \label{app:bdry_soln_1}
\end{equation}
where,
\begin{equation*}
\mathbf{G}_{\mathrm{BC}} = \left(\begin{array}{c}
G^+_\omega\left(0, \mu_i, \phi_j\right)\\
G^-_\omega\left(d, -\mu_i, \phi_j\right)
\end{array}\right)_{[2N\times 1]}
\end{equation*}
\begin{equation*}
[A]^{-1} = \left[\begin{array}{cc}
T^{+}_{kk'} & T^{-}_{kk'}\\
B^{+}_{kk'} & B^{-}_{kk'}
\end{array}\right]_{[2N\times 2N]}
\end{equation*}
and
\begin{equation*}
\bar{\tilde{\mathbf{c}}} \label{bdry_soln_1} = \left( \begin{array}{c}
\bar{c}^+_\omega\left(0, \mu_{i'}, \phi_{j'}\right)\\
\bar{c}^-_\omega\left(d, \mu_{i'}, \phi_{j'}\right)
\end{array}\right)_{[2N\times 1]}
\end{equation*}
Here, $k$ is the index for the combination $\{\mu_i, \phi_j\}$, $N$ is the total number of combinations of $\{\mu_i, \phi_j\}$ and 
\begin{align}
\label{app:c_defs}
\begin{split}
\bar{c}^+_\omega\left(0, \mu_i', \phi_j'\right) &= \frac{ p_\omega }{4\pi\mu_i'\Lambda_\omega}I^+_{\mu_i'\phi_j'} + \left(1- p_\omega \right)\left(\frac{ \sigma_\omega }{4\pi}C_\omega\Delta \bar{T}\left(z=0\right) + \frac{\left(1- \sigma_\omega \right)}{4\pi^2\Lambda_\omega}\sum_{i''j''}w_{\mu_{i''}}w_{\phi_{j''}}I^+_{i''j''}\right)\\
\bar{c}^-_\omega\left(d, \mu_i', \phi_j'\right) &= \frac{p_\omega}{4\pi\mu_i'\Lambda_\omega}I^-_{\mu_i'\phi_j'} + \left(1- p_\omega \right)\left(\frac{ \sigma_\omega }{4\pi}C_\omega\Delta \bar{T}\left(z=d\right) + \frac{\left(1- \sigma_\omega \right)}{4\pi^2\Lambda_\omega}\sum_{i''j''}w_{\mu_{i''}}w_{\phi_{j''}}I^-_{i''j''}\right)\\
\end{split}
\end{align}
With the substitution of equation~\ref{app:bdry_soln_1}, the general BTE solution (equation 11 in the main article) becomes, \\
\begin{align}
\label{app:gen_soln_1}
\begin{split}
G^+_\omega\left(z, \mu_i, \phi_j\right) &= \left(\sum_{i'j'}\left[T^{+}_{kk'}\bar{c}^+_\omega\left(0, \mu_{i'}, \phi_{j'}\right) + T^{-}_{kk'}\bar{c}^-_\omega\left(d, \mu_{i'}, \phi_{j'}\right)\right]\right)\exp\left(-\frac{\gamma^{\mathrm{FS}}_{ij}}{\mu_i\Lambda_\omega}z\right) \\
&\ \ \ \ \ +\ \frac{\exp\left(-\frac{\gamma^{\mathrm{FS}}_{ij}}{\mu_i\Lambda_\omega}z\right)}{4\pi\mu_i\Lambda_\omega}\int_0^z\left(C_\omega\Delta\bar{T} + \bar{Q}_\omega\tau_\omega\right)\exp\left(\frac{\gamma^{\mathrm{FS}}_{ij}}{\mu_i\Lambda_\omega}z'\right)\mathrm{d}z'\\
&= \left(\sum_{i'j'}\left[T^{+}_{kk'}\bar{c}^+_\omega\left(0, \mu_{i'}, \phi_{j'}\right) + T^{-}_{kk'}\bar{c}^-_\omega\left(d, \mu_{i'}, \phi_{j'}\right)\right]\right)\exp\left(-\frac{\gamma^{\mathrm{FS}}_{ij}}{\mu_i\Lambda_\omega}z\right) \\
&\ \ \ \ \ +\ \frac{1}{4\pi\mu_i\Lambda_\omega}\int_0^z\left(C_\omega\Delta\bar{T} + \bar{Q}_\omega\tau_\omega\right)\exp\left(-\frac{\gamma^{\mathrm{FS}}_{ij}}{\mu_i\Lambda_\omega}\left|z' - z\right|\right)\mathrm{d}z'\\
G^-_\omega\left(z, -\mu_i, \phi_j\right) &= \left(\sum_{i'j'}\left[B^{+}_{kk'}\bar{c}^+_\omega\left(0, \mu_{i'}, \phi_{j'}\right) + B^{-}_{kk'}\bar{c}^-_\omega\left(d, \mu_{i'}, \phi_{j'}\right)\right]\right)\exp\left(-\frac{\gamma^{\mathrm{FS}}_{ij}}{\mu_i\Lambda_\omega}\left(d-z\right)\right)\\
&\ \ \ \ \ + \frac{\exp\left(\frac{\gamma^{\mathrm{FS}}_{ij}}{\mu_i\Lambda_\omega}z\right)}{4\pi\mu_i\Lambda_\omega}\int_z^d\left(C_\omega\Delta\bar{T} + \bar{Q}_\omega\tau_\omega\right)\exp\left(-\frac{\gamma^{\mathrm{FS}}_{ij}}{\mu_i\Lambda_\omega}z'\right)\mathrm{d}z'\\
&= \left(\sum_{i'j'}\left[B^{+}_{kk'}\bar{c}^+_\omega\left(0, \mu_{i'}, \phi_{j'}\right) + B^{-}_{kk'}\bar{c}^-_\omega\left(d, \mu_{i'}, \phi_{j'}\right)\right]\right)\exp\left(-\frac{\gamma^{\mathrm{FS}}_{ij}}{\mu_i\Lambda_\omega}\left(d-z\right)\right) \\
&\ \ \ \ \ + \frac{1}{4\pi\mu_i\Lambda_\omega}\int_z^d\left(C_\omega\Delta\bar{T} + \bar{Q}_\omega\tau_\omega\right)\exp\left(-\frac{\gamma^{\mathrm{FS}}_{ij}}{\mu_i\Lambda_\omega}\left|z'-z\right|\right)\mathrm{d}z'\\
\end{split}
\end{align}
where the unknown quantities $G^+_\omega\left(z, \mu_i, \phi_j\right)$, $G^-_\omega\left(z, -\mu_i, \phi_j\right)$ and $\Delta\bar{T}$ are related through the energy conservation requirement.
\subsection{Formulation of the Integral Equation for $\Delta\bar{T}$} \label{appendix_fredholm}
To solve for the unknown quantities ($G^+_\omega\left(z, \mu_i, \phi_j\right)$, $G^-_\omega\left(z, -\mu_i, \phi_j\right)$ and $\Delta\bar{T}$), the energy conservation equation is first discretized in the angular variables ($\mu$ and $\phi$) using Gauss quadrature (equation~\ref{app:gquad}). Next, the general solution (equation~\ref{app:gen_soln_1}) is substituted into the discretized energy conservation equation to obtain the following integral equation for $\Delta\bar{T}$:
\begin{align}
\label{app:integral_1}
\begin{split}
\Delta\bar{T}\left(z\right) &= \frac{1}{\int_{\omega = 0}^{\omega_m}\frac{C_\omega}{\tau_\omega}\mathrm{d}\omega }\int_{\omega = 0}^{\omega_m}\left[\frac{1}{\tau_\omega}\sum_{ij}\left(G^+_\omega\left(z, \mu_i, \phi_j\right) + G^-_\omega\left(z, -\mu_i, \phi_j\right)\right)w_{\mu_i}w_{\phi_j}\right]\mathrm{d}\omega\\
&= \frac{1}{\int_{\omega = 0}^{\omega_m}\frac{C_\omega}{\tau_\omega}\mathrm{d}\omega } \int_{\omega = 0}^{\omega_m}\frac{1}{\tau_\omega}\Bigg[\sum_{ij}\left(\int_0^d\left(C_\omega\Delta\bar{T} + \bar{Q}_\omega\tau_\omega\right)\exp\left(-\frac{\gamma^{\mathrm{FS}}_{ij}}{\mu_i\Lambda_\omega}\left|z'-z\right|\right)\mathrm{d}z'\right)\frac{w_{\mu_i}w_{\phi_j}}{4\pi\mu_i\Lambda_\omega}\\
&\ \ \ \ \ + \sum_{ij}\sum_{i'j'}\left(T^{+}_{kk'}\bar{c}^+_\omega\left(0, \mu_{i'}, \phi_{j'}\right) + T^{-}_{kk'}\bar{c}^-_\omega\left(d, \mu_{i'}, \phi_{j'}\right)\right) w_{\mu_i}w_{\phi_j}\exp\left(-\frac{\gamma^{\mathrm{FS}}_{ij}}{\mu_i\Lambda_\omega}z\right)\\
&\ \ \ \ \ +\sum_{ij}\sum_{i'j'}\left(B^{+}_{kk'}\bar{c}^+_\omega\left(0, \mu_{i'}, \phi_{j'}\right) + B^{-}_{kk'}\bar{c}^-_\omega\left(d, \mu_{i'}, \phi_{j'}\right)\right)w_{\mu_i}w_{\phi_j}\exp\left(-\frac{\gamma^{\mathrm{FS}}_{ij}}{\mu_i\Lambda_\omega}\left(d-z\right)\right)\Bigg]\mathrm{d}\omega\\
\end{split}
\end{align}
Let us analyze the right hand side (RHS) of this equation term-by-term. For simplicity, let $\Omega = \int_{\omega = 0}^{\omega_m}\frac{C_\omega}{\tau_\omega}\mathrm{d}\omega$. The $1^{\mathrm{st}}$ term in the RHS of equation~\ref{app:integral_1} becomes:
\begin{align}
\label{app:RHS_1}
\begin{split}
& \frac{1}{\Omega}\int_{\omega = 0}^{\omega_m}\frac{1}{\tau_\omega}\left[\sum_{ij}\left(\int_0^d\left(C_\omega\Delta\bar{T} + \bar{Q}_\omega\tau_\omega\right)\exp\left(-\frac{\gamma^{\mathrm{FS}}_{ij}}{\mu_i\Lambda_\omega}\left|z'-z\right|\right)\mathrm{d}z'\right)\frac{w_{\mu_i}w_{\phi_j}}{4\pi\mu_i\Lambda_\omega}\mathrm{d}\omega\right] \\
&= \frac{1}{\Omega}\int_0^d\Delta\bar{T}\left[\int_{\omega=0}^{\omega_m}\left(\frac{C_\omega}{4\pi\tau_\omega\Lambda_\omega}\sum_{ij}\frac{w_{\mu_i}w_{\phi_j}}{\mu_i}\exp\left(-\frac{\gamma^{\mathrm{FS}}_{ij}}{\mu_i\Lambda_\omega}\left|z'-z\right|\right)\right)\mathrm{d}\omega\right]\mathrm{d}z'\\
&\ \ \ \ \ +\frac{1}{\Omega}\int_{\omega = 0}^{\omega_m}\bar{Q}_\omega\left[\sum_{ij}\left(\int_0^z \exp\left(-\frac{\gamma^{\mathrm{FS}}_{ij}}{\mu_i\Lambda_\omega}\left|z'-z\right|\right)\mathrm{d}z' + \int_z^d \exp\left(-\frac{\gamma^{\mathrm{FS}}_{ij}}{\mu_i\Lambda_\omega}\left|z'-z\right|\right)\mathrm{d}z'\right)\frac{w_{\mu_i}w_{\phi_j}}{4\pi\mu_i\Lambda_\omega}\right]\\
&= \frac{1}{\Omega}\int_0^d\Delta\bar{T}\left[\int_{\omega=0}^{\omega_m}\left(\frac{C_\omega}{4\pi\tau_\omega\Lambda_\omega}\sum_{ij}\frac{w_{\mu_i}w_{\phi_j}}{\mu_i}\exp\left(-\frac{\gamma^{\mathrm{FS}}_{ij}}{\mu_i\Lambda_\omega}\left|z'-z\right|\right)\right)\mathrm{d}\omega\right]\mathrm{d}z'\\
&\ \ \ \ \ +\frac{1}{\Omega}\int_{\omega = 0}^{\omega_m}\bar{Q}_\omega\left[\sum_{ij}\left(2-\exp\left(-\frac{{\gamma^{\mathrm{FS}}_{ij}}}{\mu_i\Lambda_\omega}z\right) -\exp\left(-\frac{{\gamma^{\mathrm{FS}}_{ij}}}{\mu_i\Lambda_\omega}\left(d-z\right)\right)\right)\frac{w_{\mu_i}w_{\phi_j}}{4\pi\gamma^{\mathrm{FS}}_{ij}}\right]\mathrm{d}\omega\\
&= \int_0^d\Delta\bar{T}\left[K^1_1\left(z', z\right)\right]\mathrm{d}z' + f^1_1\left(z\right)
\end{split}
\end{align}
Similarly, the $2^{\mathrm{nd}}$ term in the RHS of equation~\ref{app:integral_1} becomes:
\begin{align}
\label{app:RHS_2}
\begin{split}
& \frac{1}{\Omega}\int_{\omega=0}^{\omega_m}\frac{1}{\tau_\omega}\Bigg[\sum_{ij}\sum_{i'j'}\left(T^{+}_{kk'}\bar{c}^+_\omega\left(0, \mu_{i'}, \phi_{j'}\right) + T^{-}_{kk'}\bar{c}^-_\omega\left(d, \mu_{i'}, \phi_{j'}\right)\right) w_{\mu_i}w_{\phi_j}\exp\left(-\frac{\gamma^{\mathrm{FS}}_{ij}}{\mu_i\Lambda_\omega}z\right)\Bigg]\mathrm{d}\omega\\
&\ \ \ \ \ =\frac{1}{\Omega}\int_{\omega=0}^{\omega_m}\frac{1}{\tau_\omega}\Bigg[\sum_{ij}\sum_{i'j'}\Bigg(\frac{p_\omega}{4\pi\mu_{i'}\Lambda_\omega}\left(I^+_{i'j'}T_{kk'}^{+} + I^-_{i'j'}T_{kk'}^{-} \right)w_{\mu_i}w_{\phi_j}\exp\left(-\frac{\gamma^{\mathrm{FS}}_{ij}}{\mu_i\Lambda_\omega}z\right)\Bigg)\\ 
&\ \ \ \ \ \ \ +\left(1- p_\omega \right)\left(1- \sigma_\omega \right)\sum_{ij}\sum_{i'j'}\sum_{i''j''}\Bigg(\frac{w_{\mu_{i''}}w_{\phi_{j''}}}{4\pi ^2\Lambda_\omega}\Bigg(I^+_{i''j''}T_{kk'}^{+} +I^-_{i''j''}T_{kk'}^{-}\Bigg)w_{\mu_i}w_{\phi_j}\exp\left(-\frac{\gamma^{\mathrm{FS}}_{ij}}{\mu_i\Lambda_\omega}z\right)\Bigg) \\
&\ \ \ \ \ \ \ + \left(1- p_\omega \right) \sigma_\omega\sum_{ij}\sum_{i'j'}\Bigg(\frac{C_\omega}{4\pi}\Bigg(T_{kk'}^{+} \Delta \bar{T}\left(z=0\right) + T_{kk'}^{-} \Delta \bar{T}\left(z=d\right)\Bigg)w_{\mu_i}w_{\phi_j}\exp\left(-\frac{\gamma^{\mathrm{FS}}_{ij}}{\mu_i\Lambda_\omega}z\right)\Bigg)\Bigg]\mathrm{d}\omega\\
&\ \ \ \ \ = f^1_2\left(z\right)+ f^2_2\left(z\right)+ h_2\left(z\right) + \int_0^d\Delta\bar{T}\left[K^1_2\left(z', z\right) + K^2_2\left(z', z\right)\right]\mathrm{d}z'
\end{split}
\end{align}
where,
\begin{align}
\label{app:f12}
\begin{split}
f^1_2\left(z\right) &= \frac{1}{\Omega}\int_{\omega=0}^{\omega_m}\Bigg[\sum_{ij}\sum_{i'j'}\Bigg(\frac{\bar{Q}_\omega p_\omega\left(T_{kk'}^{+} + T_{kk'}^{-} \right)}{4\pi\gamma^{\mathrm{FS}}_{\mu_i'\phi_j'}}\left(1-\exp\left(-\frac{\gamma^{\mathrm{FS}}_{i'j'}}{\mu_{i'}\Lambda_\omega}d\right)\right)\exp\left(-\frac{\gamma^{\mathrm{FS}}_{ij}}{\mu_i\Lambda_\omega}z\right)w_{\mu_i}w_{\phi_j}\Bigg)\Bigg]\mathrm{d}\omega
\end{split}
\end{align}
\begin{align}
\label{app:f22}
\begin{split}
f^2_2\left(z\right) &= \frac{1}{\Omega}\int_{\omega = 0}^{\omega_m}\Bigg[\sum_{ij}\sum_{i'j'}\sum_{i''j''}\Bigg(\frac{\bar{Q}_\omega\left(1- p_\omega \right)\left(1- \sigma_\omega \right)\left(T_{kk'}^{+} + T_{kk'}^{-}\right) \mu_{i''}w_{\mu_{i''}}w_{\phi_{k''}}}{4\pi^2\gamma^{\mathrm{FS}}_{i''j''}}\\
&\ \ \ \ \ \times\left(1-\exp\left(-\frac{\gamma^{\mathrm{FS}}_{i''j''}}{\mu_{i''}\Lambda_\omega}d\right)\right) w_{\mu_i}w_{\phi_j}\exp\left(-\frac{\gamma^{\mathrm{FS}}_{ij}}{\mu_i\Lambda_\omega}z\right)\Bigg)\Bigg]\mathrm{d}\omega\\
\end{split}
\end{align}
\begin{align}
\label{app:h2}
\begin{split}
h_2\left(z\right) &= \frac{1}{\Omega}\int_{\omega=0}^{\omega_m}\frac{1}{\tau_\omega}\Bigg[\sum_{ij}\sum_{i'j'}\Bigg(\frac{C_\omega\left(1- p_\omega \right) \sigma_\omega}{4\pi}\Bigg(T_{kk'}^{+} \Delta \bar{T}\left(z=0\right) + T_{kk'}^{-}\Delta \bar{T}\left(z=d\right)\Bigg)\\
&\ \ \ \ \ \times w_{\mu_i}w_{\phi_j}\exp\left(-\frac{\gamma^{\mathrm{FS}}_{ij}}{\mu_i\Lambda_\omega}z\right)\Bigg)\Bigg]
\end{split}
\end{align}
\begin{align}
\label{app:K12}
\begin{split}
K^1_2\left(z', z\right) &= \frac{1}{\Omega}\int_{\omega=0}^{\omega_m}\frac{C_\omega}{\tau_\omega}\Bigg[\sum_{ij}\sum_{i'j'}\Bigg(\frac{p_\omega}{4\pi\mu_{i'}\Lambda_\omega}\Bigg(\exp\left(-\frac{\gamma^{\mathrm{FS}}_{i'j'}}{\mu_{i'}\Lambda_\omega}z'\right)T_{kk'}^{+} \\
&\ \ \ \ \ + \exp\left(-\frac{\gamma^{\mathrm{FS}}_{i'j'}}{\mu_{i'}\Lambda_\omega}\left(d-z'\right)\right)T_{kk'}^{-} \Bigg)w_{\mu_i}w_{\phi_j}\exp\left(-\frac{\gamma^{\mathrm{FS}}_{ij}}{\mu_i\Lambda_\omega}z\right)\Bigg)\Bigg]\mathrm{d}\omega
\end{split}
\end{align}
\begin{align}
\label{app:K22}
\begin{split}
K^2_2\left(z', z\right) &= \frac{1}{\Omega}\int_{\omega=0}^{\omega_m}\frac{C_\omega\left(1- p_\omega \right)\left(1- \sigma_\omega \right)}{\tau_\omega}\Bigg[\sum_{ij}\sum_{i'j'}\sum_{i''j''}\Bigg(\frac{w_{\mu_{i''}}w_{\phi_{j''}}}{4\pi ^2\Lambda_\omega}\Bigg(\exp\left(-\frac{\gamma^{\mathrm{FS}}_{i''j''}}{\mu_{i''}\Lambda_\omega}z'\right)T_{kk'}^{+}\\
&\ \ \ \ \ \ \ \ \ + \exp\left(-\frac{\gamma^{\mathrm{FS}}_{i''j''}}{\mu_{i''}\Lambda_\omega}\left(d-z'\right)\right)T_{kk'}^{-}\Bigg) w_{\mu_i}w_{\phi_j}\exp\left(-\frac{\gamma^{\mathrm{FS}}_{ij}}{\mu_i\Lambda_\omega}z\right)\Bigg)\Bigg]\mathrm{d}\omega\\
\end{split}
\end{align}
and the $3^{\mathrm{rd}}$ term in the RHS of equation~\ref{app:integral_1} becomes
\begin{align}
\label{app:RHS_3}
\begin{split}
& \frac{1}{\Omega}\int_{\omega=0}^{\omega_m}\frac{1}{\tau_\omega}\Bigg[\sum_{ij}\sum_{i'j'}\left(B^{+}_{kk'}\bar{c}^+_\omega\left(0, \mu_{i'}, \phi_{j'}\right) + B^{-}_{kk'}\bar{c}^-_\omega\left(d, \mu_{i'}, \phi_{j'}\right)\right) w_{\mu_i}w_{\phi_j}\exp\left(-\frac{\gamma^{\mathrm{FS}}_{ij}}{\mu_i\Lambda_\omega}\left(d-z\right)\right)\Bigg]\mathrm{d}\omega\\
&\ \ \ \ \ =\frac{1}{\Omega}\int_{\omega=0}^{\omega_m}\frac{1}{\tau_\omega}\Bigg[\sum_{ij}\sum_{i'j'}\Bigg(\frac{p_\omega}{4\pi\mu_{i'}\Lambda_\omega}\left(I^+_{i'j'}B_{kk'}^{+} + I^-_{i'j'}B_{kk'}^{-} \right)w_{\mu_i}w_{\phi_j}\exp\left(-\frac{\gamma^{\mathrm{FS}}_{ij}}{\mu_i\Lambda_\omega}\left(d-z\right)\right)\Bigg)\\ 
&\ \ \ \ \ \ \ +\sum_{ij}\sum_{i'j'}\sum_{i''j''}\Bigg(\frac{w_{\mu_{i''}}w_{\phi_{j''}}\left(1- p_\omega \right)\left(1- \sigma_\omega \right)}{4\pi ^2\Lambda_\omega}\Bigg(I^+_{i''j''}B_{kk'}^{+} +I^-_{i''j''}B_{kk'}^{-}\Bigg) w_{\mu_i}w_{\phi_j}\exp\left(-\frac{\gamma^{\mathrm{FS}}_{ij}}{\mu_i\Lambda_\omega}\left(d-z\right)\right)\Bigg)\\
&\ \ \ \ \ \ \ + \sum_{ij}\sum_{i'j'}\Bigg(\frac{C_\omega\left(1- p_\omega \right) \sigma_\omega}{4\pi}\Bigg(B_{kk'}^{+} \Delta \bar{T}\left(z=0\right)+ B_{kk'}^{-} \Delta \bar{T}\left(z=d\right)\Bigg)w_{\mu_i}w_{\phi_j}\exp\left(-\frac{\gamma^{\mathrm{FS}}_{ij}}{\mu_i\Lambda_\omega}\left(d-z\right)\right)\Bigg)\Bigg]\mathrm{d}\omega\\
&\ \ \ \ \ = f^1_3\left(z\right)+ f^2_3\left(z\right)+ h_3\left(z\right) + \int_0^d\Delta\bar{T}\left[K^1_3\left(z', z\right) + K^2_3\left(z', z\right)\right]\mathrm{d}z'
\end{split}
\end{align}
where,
\begin{align}
\label{app:f13}
\begin{split}
f^1_3\left(z\right) &= \frac{1}{\Omega}\int_{\omega=0}^{\omega_m}\Bigg[\sum_{ij}\sum_{i'j'}\Bigg(\frac{\bar{Q}_\omega p_\omega}{4\pi\gamma^{\mathrm{FS}}_{\mu_i'\phi_j'}}\left(1-\exp\left(-\frac{\gamma^{\mathrm{FS}}_{i'j'}}{\mu_{i'}\Lambda_\omega}d\right)\right)\Bigg(B_{kk'}^{+} + B_{kk'}^{-} \Bigg)\\
&\ \ \ \ \ w_{\mu_i}w_{\phi_j}\exp\left(-\frac{\gamma^{\mathrm{FS}}_{ij}}{\mu_i\Lambda_\omega}\left(d-z\right)\right)\Bigg)\Bigg]\mathrm{d}\omega
\end{split}
\end{align}
\begin{align}
\label{app:f23}
\begin{split}
f^2_3\left(z\right) &= \frac{1}{\Omega}\int_{\omega = 0}^{\omega_m}\Bigg[\sum_{ij}\sum_{i'j'}\sum_{i''j''}\Bigg(\frac{\bar{Q}_\omega \mu_{i''}w_{\mu_{i''}}w_{\phi_{k''}}\left(1- p_\omega \right)\left(1- \sigma_\omega \right)}{4\pi^2\gamma^{\mathrm{FS}}_{i''j''}}\left(1-\exp\left(-\frac{\gamma^{\mathrm{FS}}_{i''j''}}{\mu_{i''}\Lambda_\omega}d\right)\right)\\
&\ \ \ \ \ \ \times\left(B_{kk'}^{+} + B_{kk'}^{-}\right) w_{\mu_i}w_{\phi_j}\exp\left(-\frac{\gamma^{\mathrm{FS}}_{ij}}{\mu_i\Lambda_\omega}\left(d-z\right)\right)\Bigg)\Bigg]\mathrm{d}\omega\\
\end{split}
\end{align}
\begin{align}
\label{app:h3}
\begin{split}
h_3\left(z\right) &= \frac{1}{\Omega}\int_{\omega=0}^{\omega_m}\frac{1}{\tau_\omega}\Bigg[\sum_{ij}\sum_{i'j'}\Bigg(\frac{C_\omega\left(1- p_\omega \right) \sigma_\omega}{4\pi}\Bigg(B_{kk'}^{+} \Delta \bar{T}\left(z=0\right) \\
&\ \ \ \ \ + B_{kk'}^{-} \Delta \bar{T}\left(z=d\right)\Bigg)w_{\mu_i}w_{\phi_j}\exp\left(-\frac{\gamma^{\mathrm{FS}}_{ij}}{\mu_i\Lambda_\omega}\left(d-z\right)\right)\Bigg)\Bigg]
\end{split}
\end{align}
\begin{align}
\label{app:K13}
\begin{split}
K^1_3\left(z', z\right) &= \frac{1}{\Omega}\int_{\omega=0}^{\omega_m}\frac{C_\omega}{\tau_\omega}\Bigg[\sum_{ij}\sum_{i'j'}\Bigg(\frac{p_\omega}{4\pi\mu_{i'}\Lambda_\omega}\Bigg(\exp\left(-\frac{\gamma^{\mathrm{FS}}_{i'j'}}{\mu_{i'}\Lambda_\omega}z'\right)B_{kk'}^{+} \\
&\ \ \ \ \ + \exp\left(-\frac{\gamma^{\mathrm{FS}}_{i'j'}}{\mu_{i'}\Lambda_\omega}\left(d-z'\right)\right)B_{kk'}^{-} \Bigg) w_{\mu_i}w_{\phi_j}\exp\left(-\frac{\gamma^{\mathrm{FS}}_{ij}}{\mu_i\Lambda_\omega}\left(d-z\right)\right)\Bigg)\Bigg]\mathrm{d}\omega
\end{split}
\end{align}
\begin{align}
\label{app:K23}
\begin{split}
K^2_3\left(z', z\right) &= \frac{1}{\Omega}\int_{\omega=0}^{\omega_m}\frac{C_\omega}{\tau_\omega}\Bigg[\sum_{ij}\sum_{i'j'}\sum_{i''j''}\Bigg(\frac{w_{\mu_{i''}}w_{\phi_{j''}}\left(1- p_\omega \right)\left(1- \sigma_\omega \right)}{4\pi ^2\Lambda_\omega}\Bigg(\exp\left(-\frac{\gamma^{\mathrm{FS}}_{i''j''}}{\mu_{i''}\Lambda_\omega}z'\right)B_{kk'}^{+} \\
&\ \ \ \ \ \ \ \ \ \ + \exp\left(-\frac{\gamma^{\mathrm{FS}}_{i''j''}}{\mu_{i''}\Lambda_\omega}\left(d-z'\right)\right)B_{kk'}^{-}\Bigg)w_{\mu_i}w_{\phi_j}\exp\left(-\frac{\gamma^{\mathrm{FS}}_{ij}}{\mu_i\Lambda_\omega}\left(d-z\right)\right)\Bigg)\Bigg]\mathrm{d}\omega\\
\end{split}
\end{align}
Finally, the system to solve for (equation~\ref{app:integral_1}) can be represented as an integral equation of the form:
\begin{equation}
\Delta\bar{T}\left(z\right) - h\left(z\right) = f\left(z\right) + \int_0^d\left[K\left(z', z\right)\Delta\bar{T}\left(z'\right)\right]\mathrm{d}z' \label{app:integral_2}
\end{equation}
where,
\begin{equation}
f\left(z\right) = f^1_1\left(z\right) + f^1_2\left(z\right) + f^2_2\left(z\right) + f^1_3\left(z\right) + f^2_3\left(z\right) \label{app:f_def}
\end{equation}
\begin{equation}
K\left(z', z\right) = K^1_1\left(z', z\right) + K^1_2\left(z', z\right) + K^2_2\left(z', z\right) + K^1_3\left(z', z\right) + K^2_3\left(z', z\right) \label{app:k_def}
\end{equation}
and
\begin{equation}
h\left(z\right) = h_2\left(z\right) + h_3\left(z\right) \label{app:h_def}
\end{equation}
There are several important properties of the integral equation~\ref{app:integral_2}. 
\begin{enumerate}
\item The kernel $K\left(z', z\right)$ is singular for $z=z'$ due to the singularity of $K_1^1\left(z', z\right)$ at $z=z'$. 
\item Unlike the term $f\left(z\right)$, the term $h\left(z\right)$ is a function of $\Delta\bar{T}$ and can be represented as $h\left(z\right) = H\left(z', z\right)\Delta\bar{T}$ where $H\left(z', z\right)$ is independent of $\Delta\bar{T}$.
\end{enumerate}
There are several approaches available in the literature to solve such singular integral equations. In this work, this integral equation is solved using the method of degenerate kernels, the details of which are described in section~\ref{appendix_degenerate_kernels}.
\subsection{The Method of Degenerate Kernels} \label{appendix_degenerate_kernels}
The integral equation (equation~\ref{app:integral_2}) can be solved using the method of degenerate kernels. First, the integral equation is rewritten as,
\begin{equation}
\Delta\bar{T}\left(\hat{z}\right) - h\left(\hat{z}\right) = f\left(\hat{z}\right) + \int_0^1\left[\bar{K}\left(\hat{z}', \hat{z}\right)\Delta\bar{T}\left(\hat{z}'\right)\right]\mathrm{d}\hat{z}' \label{app:integral_3}
\end{equation}
where $\hat{z} = z/d$ and $\bar{K}\left(\hat{z}', \hat{z}\right) = d\times K\left(z', z\right)$. Then the functions $\Delta\bar{T}\left(\hat{z}\right)$, $f\left(\hat{z}\right)$ and $K\left(\hat{z}', \hat{z}\right)$ are expanded in a Fourier series :
\begin{equation}
\Delta\bar{T}_{\left(N\right)}\left(\hat{z}\right) = \frac{1}{2}t_0 + \sum_{m=1}^Nt_m\cos\left(m\pi\hat{z}\right) \label{app:T_Fourier}
\end{equation}
\begin{equation}
f_{\left(N\right)}\left(\hat{z}\right) = \frac{1}{2}f_0 + \sum_{m=1}^Nf_n\cos\left(m\pi\hat{z}\right) \label{app:f_Fourier}
\end{equation}
\begin{equation}
h_{\left(N\right)}\left(\hat{z}\right) = \frac{1}{2}h_0 + \sum_{m=1}^Nh_n\cos\left(m\pi\hat{z}\right) \label{app:f_Fourier}
\end{equation}
\begin{align*}
\bar{K}_{\left(N\right)}\left(\hat{z}', \hat{z}\right) &= \frac{1}{4}K_{00} + \frac{1}{2}\sum_{m=1}^NK_{m0}\cos\left(m\pi\hat{z}'\right) + \frac{1}{2}\sum_{n=1}^NK_{0n}\cos\left(n\pi\hat{z}\right) \\
&\ \ \ + \sum_{m=1}^N\sum_{n=1}^NK_{mn}\cos\left(m\pi\hat{z}'\right)\cos\left(n\pi\hat{z}\right) \label{app:K_Fourier}
\end{align*}
where the Fourier coefficients are given by,
\begin{equation}
f_m = 2\int_0^1f\left(\hat{z}\right)\cos\left(m\pi\hat{z}\right)\mathrm{d}\hat{z} \label{app:f_Fourier_coeff}
\end{equation}
\begin{equation}
h_m = 2\int_0^1h\left(\hat{z}\right)\cos\left(m\pi\hat{z}\right)\mathrm{d}\hat{z} \label{app:h_Fourier_coeff}
\end{equation}
and
\begin{equation}
K_{mn} = 4\int_0^1\int_0^1K\left(\hat{z}', \hat{z}\right)\cos\left(m\pi\hat{z}'\right)\cos\left(n\pi\hat{z}\right)\mathrm{d}\hat{z}'\mathrm{d}\hat{z} \label{app:K_Fourier_coeff}
\end{equation}
Here, a Fourier cosine series has been used for all of the functions by assuming that all the functions are even with respect to $\hat{z}$ and $\hat{z}'$. This assumption is valid since the integral equation (equation~\ref{app:integral_2}) has been solved only in the domain $\hat{z}\in\left[0, 1\right]$. After several algebraic simplifications, the expressions for the Fourier coefficients (equations~\ref{app:f_Fourier_coeff}, ~\ref{app:h_Fourier_coeff} and~\ref{app:K_Fourier_coeff}) simplify into the following concise forms:
\begin{align}
\label{app:f11_m}
\begin{split}
f^1_{1, m} &= -\frac{2}{\Omega}\int_{\omega = 0}^{\omega_m}\bar{Q}_\omega\left[\sum_{ij}\left(I_1\left(m\right) + I_2\left(m\right)\right)\frac{w_{\mu_i}w_{\phi_j}}{4\pi\gamma^{\mathrm{FS}}_{ij}}\right]\mathrm{d}\omega
\end{split}
\end{align}
\begin{align}
\label{app:f12_m}
\begin{split}
f^1_{2, m} &= \frac{2}{\Omega}\int_{\omega=0}^{\omega_m}\Bigg[\sum_{ij}\sum_{i'j'}\Bigg(\frac{\bar{Q}_\omega p_\omega}{4\pi\gamma^{\mathrm{FS}}_{\mu_i'\phi_j'}}\left(1-\exp\left(-\frac{\gamma^{\mathrm{FS}}_{i'j'}}{\mu_{i'}\Lambda_\omega}d\right)\right)\left(T_{kk'}^{+} + T_{kk'}^{-} \right)w_{\mu_i}w_{\phi_j}I_1\left(m\right)\Bigg)\Bigg]\mathrm{d}\omega
\end{split}
\end{align}
\begin{align}
\label{app:f22_m}
\begin{split}
f^2_{2, m} &=\frac{2}{\Omega}\int_{\omega = 0}^{\omega_m}\Bigg[\sum_{ij}\sum_{i'j'}\sum_{i''j''}\Bigg(\frac{\bar{Q}_\omega \mu_{i''}w_{\mu_{i''}}w_{\phi_{k''}}\left(1- p_\omega \right)\left(1- \sigma_\omega \right)}{4\pi^2\gamma^{\mathrm{FS}}_{i''j''}}\left(1-\exp\left(-\frac{\gamma^{\mathrm{FS}}_{i''j''}}{\mu_{i''}\Lambda_\omega}d\right)\right)\\
&\ \ \ \ \ \times\left(T_{kk'}^{+} + T_{kk'}^{-}\right) w_{\mu_i}w_{\phi_j}I_1\left(m\right)\Bigg)\Bigg]\mathrm{d}\omega
\end{split}
\end{align}
\begin{align}
\label{app:h2_m}
\begin{split}
h_{2, m} &= \frac{2}{\Omega}\int_{\omega=0}^{\omega_m}\frac{1}{\tau_\omega}\Bigg[\sum_{ij}\sum_{i'j'}\Bigg(\frac{C_\omega\left(1- p_\omega \right) \sigma_\omega}{4\pi}\Bigg(T_{kk'}^{+} \Delta \bar{T}\left(z=0\right) + T_{kk'}^{-}\Delta \bar{T}\left(z=d\right)\Bigg)w_{\mu_i}w_{\phi_j}I_1\left(m\right)\Bigg)\Bigg]
\end{split}
\end{align}
\begin{align}
\label{app:f13_m}
\begin{split}
f^1_{3, m} &= \frac{2}{\Omega}\int_{\omega=0}^{\omega_m}\Bigg[\sum_{ij}\sum_{i'j'}\Bigg(\frac{\bar{Q}_\omega p_\omega}{4\pi\gamma^{\mathrm{FS}}_{\mu_i'\phi_j'}}\left(1-\exp\left(-\frac{\gamma^{\mathrm{FS}}_{i'j'}}{\mu_{i'}\Lambda_\omega}d\right)\right)\left(B_{kk'}^{+} + B_{kk'}^{-} \right)w_{\mu_i}w_{\phi_j}I_2\left(m\right)\Bigg)\Bigg]\mathrm{d}\omega 
\end{split}
\end{align}
\begin{align}
\label{app:f23_m}
\begin{split}
f^2_{3, m} &= \frac{2}{\Omega}\int_{\omega = 0}^{\omega_m}\Bigg[\sum_{ij}\sum_{i'j'}\sum_{i''j''}\Bigg(\frac{\bar{Q}_\omega\left(1- p_\omega \right)\left(1- \sigma_\omega \right) \mu_{i''}w_{\mu_{i''}}w_{\phi_{k''}}}{4\pi^2\gamma^{\mathrm{FS}}_{i''j''}}\left(1-\exp\left(-\frac{\gamma^{\mathrm{FS}}_{i''j''}}{\mu_{i''}\Lambda_\omega}d\right)\right)\\
& \ \ \ \ \ \times\left(B_{kk'}^{+} + B_{kk'}^{-}\right) w_{\mu_i}w_{\phi_j}I_2\left(m\right)\Bigg)\Bigg]\mathrm{d}\omega
\end{split}
\end{align}
\begin{align}
\label{app:h3_m}
\begin{split}
h_{3, m} &= \frac{2}{\Omega}\int_{\omega=0}^{\omega_m}\frac{1}{\tau_\omega}\Bigg[\sum_{ij}\sum_{i'j'}\Bigg(\frac{C_\omega\left(1- p_\omega \right) \sigma_\omega}{4\pi}\Bigg(B_{kk'}^{+} \Delta \bar{T}\left(z=0\right) + B_{kk'}^{-}\Delta \bar{T}\left(z=d\right)\Bigg)w_{\mu_i}w_{\phi_j}I_2\left(m\right)\Bigg)\Bigg]
\end{split}
\end{align}
\begin{align}
\label{app:K11_mn}
\begin{split}
K^1_{1, mn} &= \frac{4d}{\Omega}\left[\int_{\omega=0}^{\omega_m}\left(\frac{C_\omega}{4\pi\tau_\omega\Lambda_\omega}\sum_{ij}\frac{w_{\mu_i}w_{\phi_j}}{\mu_i}I_3\left(m, n\right)\right)\mathrm{d}\omega\right]\\
\end{split}
\end{align}
\begin{align}
\label{app:K12_mn}
\begin{split}
K^1_{2, mn} &= \frac{4d}{\Omega}\int_{\omega=0}^{\omega_m}\frac{C_\omega}{\tau_\omega}\Bigg[\sum_{ij}\sum_{i'j'}\Bigg(\frac{p_\omega}{4\pi\mu_{i'}\Lambda_\omega}\left(I_1'\left(m\right)T_{kk'}^{+} + I_2'\left(m\right)T_{kk'}^{-} \right)w_{\mu_i}w_{\phi_j}I_1\left(n\right)\Bigg)\Bigg]\mathrm{d}\omega\\
\end{split}
\end{align}
\begin{align}
\label{app:K22_mn}
\begin{split}
K^2_{2, mn} &= \frac{4d}{\Omega}\int_{\omega=0}^{\omega_m}\frac{C_\omega}{\tau_\omega}\Bigg[\sum_{ij}\sum_{i'j'}\sum_{i''j''}\Bigg(\frac{w_{\mu_{i''}}w_{\phi_{j''}}\left(1- p_\omega \right)\left(1- \sigma_\omega \right)}{4\pi ^2\Lambda_\omega}\Bigg(I''_1\left(m\right)T_{kk'}^{+}\\
&\ \ \ \ \ + I''_2\left(m\right)T_{kk'}^{-}\Bigg)w_{\mu_i}w_{\phi_j}I_1\left(n\right)\Bigg)\Bigg]\mathrm{d}\omega
\end{split}
\end{align}
\begin{align}
\label{app:K13_mn}
\begin{split}
K^1_{3, mn} &= \frac{4d}{\Omega}\int_{\omega=0}^{\omega_m}\frac{C_\omega}{\tau_\omega}\Bigg[\sum_{ij}\sum_{i'j'}\Bigg(\frac{p_\omega}{4\pi\mu_{i'}\Lambda_\omega} \left(I'_1\left(m\right)B_{kk'}^{+} + I'_2\left(m\right)B_{kk'}^{-} \right)w_{\mu_i}w_{\phi_j}I_2\left(n\right)\Bigg)\Bigg]\mathrm{d}\omega
\end{split}
\end{align}
\begin{align}
\label{app:K23_mn}
\begin{split}
K^2_{3, mn} &= \frac{4d}{\Omega}\int_{\omega=0}^{\omega_m}\frac{C_\omega}{\tau_\omega}\Bigg[\sum_{ij}\sum_{i'j'}\sum_{i''j''}\Bigg(\frac{w_{\mu_{i''}}w_{\phi_{j''}}\left(1-p_\omega\right)\left(1- \sigma_\omega \right)}{4\pi ^2\Lambda_\omega}\Bigg(I''_1\left(m\right)B_{kk'}^{+} \\
&\ \ \ \ \ + I''_2\left(m\right)B_{kk'}^{-}\Bigg)w_{\mu_i}w_{\phi_j}I_2\left(n\right)\Bigg)\Bigg]\mathrm{d}\omega\\
\end{split}
\end{align}
where,
\begin{align*}
I_1\left(m\right) &= \int_0^1\exp\left(-\frac{{\gamma^{\mathrm{FS}}_{ij}}}{\mu_i\mathrm{Kn}^d_\omega}\hat{z}\right)\cos\left(m\pi\hat{z}\right)\mathrm{d}\hat{z}\\
&= \frac{\frac{{\gamma^{\mathrm{FS}}_{ij}}}{\mu_i\mathrm{Kn}^d_\omega}}{m^2\pi^2 + \left(\frac{{\gamma^{\mathrm{FS}}_{ij}}}{\mu_i\mathrm{Kn}^d_\omega}\right)^2}\left[1 - \left(-1\right)^m\exp\left(-\frac{{\gamma^{\mathrm{FS}}_{ij}}}{\mu_i\mathrm{Kn}^d_\omega}\right)\right]
\end{align*}
\begin{align*}
I_2\left(m\right) &= \int_0^1\exp\left(-\frac{{\gamma^{\mathrm{FS}}_{ij}}}{\mu_i\mathrm{Kn}^d_\omega}\left(1-\hat{z}\right)\right)\cos\left(m\pi\hat{z}\right)\mathrm{d}\hat{z}\\
&= \frac{\frac{{\gamma^{\mathrm{FS}}_{ij}}}{\mu_i\mathrm{Kn}^d_\omega}}{m^2\pi^2 + \left(\frac{{\gamma^{\mathrm{FS}}_{ij}}}{\mu_i\mathrm{Kn}^d_\omega}\right)^2}\left[\left(-1\right)^m - \exp\left(-\frac{{\gamma^{\mathrm{FS}}_{ij}}}{\mu_i\mathrm{Kn}^d_\omega}\right)\right]
\end{align*}
\begin{align*}
I_3\left(m, n\right) &= \int_0^1\int_0^1\exp\left(-\frac{{\gamma^{\mathrm{FS}}_{ij}}}{\mu_i\mathrm{Kn}^d_\omega}\left|\hat{z}' - \hat{z}\right|\right)\cos\left(m\pi\hat{z}\right)\cos\left(n\pi\hat{z}'\right)\mathrm{d}\hat{z}\mathrm{d}\hat{z}'\\
&= \begin{cases}
\frac{\frac{\gamma^{\mathrm{FS}}_{ij}}{\mu_i\mathrm{Kn}^d_\omega}}{m^2\pi^2 + \left(\frac{\gamma^{\mathrm{FS}}_{ij}}{\mu_i\mathrm{Kn}^d_\omega}\right)^2}\left[\delta_{mn} - \left(I_1\left(n\right) + \left(-1\right)^mI_2\left(n\right)\right)\right] & \mathrm{for}\ m\neq 0\\
\frac{\frac{\gamma^{\mathrm{FS}}_{ij}}{\mu_i\mathrm{Kn}^d_\omega}}{m^2\pi^2 + \left(\frac{\gamma^{\mathrm{FS}}_{ij}}{\mu_i\mathrm{Kn}^d_\omega}\right)^2}\left[2\delta_{mn} - \left(I_1\left(n\right) + \left(-1\right)^mI_2\left(n\right)\right)\right] & \mathrm{for}\ m = 0\\
\end{cases}
\end{align*}
and primes ($'$ and $''$) on $I_1, I_2$ and $I_3$ indicate that these functions are evaluated for $\{\mu', \phi'\}$ and $\{\mu'', \phi''\}$ respectively. These Fourier coefficients are substituted into the cosine series for the corresponding functions in the integral equation (equation~\ref{app:integral_2}) to get,
\begin{align*}
\frac{1}{2}t_0 &+ \sum_{m=1}^Nt_m\cos\left(m\pi\hat{z}\right) - \frac{1}{2}h_0 - \sum_{m=1}^Nh_m\cos\left(m\pi\hat{z}\right)\\
& = \frac{1}{2}f_0 + \sum_{m=1}^Nf_n\cos\left(m\pi\hat{z}\right) + \frac{1}{8}t_0K_{00} + \frac{1}{4}\sum_{n=1}^Nt_0K_{0n}\cos\left(n\pi\hat{z}\right) \\
&\ \ \ \ \ + \sum_{m=1}^N\left(\frac{t_0K_{m0}}{4} + \frac{K_{00}t_m}{4} + \sum_{n=1}^N\frac{t_0K_{mn} + t_mK_{0n}}{2}\cos\left(n\pi\hat{z}\right)\right)\int_0^1\cos\left(m\pi\hat{z}'\right)\mathrm{d}\hat{z}'\\
&\ \ \ \ \ + \sum_{m=1}^N\sum_{n=1}^N\left(\frac{t_mK_{n0}}{2} + \sum_{p=1}^Nt_mK_{np}\cos\left(p\pi\hat{z}\right)\right)\int_0^1\cos\left(m\pi\hat{z}'\right)\cos\left(n\pi\hat{z}'\right)\mathrm{d}\hat{z}'\\
&= \frac{1}{2}f_0 + \sum_{m=1}^Nf_n\cos\left(m\pi\hat{z}\right) +\ \frac{1}{8}t_0K_{00} + \frac{1}{4}\sum_{n=1}^Nt_0K_{0n}\cos\left(n\pi\hat{z}\right)\\
&\ \ \ \ \ +\frac{1}{2}\sum_{m=1}^N\left(\frac{t_mK_{m0}}{2} + \sum_{n=1}^Nt_mK_{mn}\cos\left(n\pi\hat{z}\right)\right)
\end{align*}
Due to the orthogonality of $\cos\left(m\pi\hat{z}\right)$ in the interval $\hat{z}\in[0, 1]$, it is sufficient to solve for the Fourier coefficients ($t_m$ ) by grouping together the coefficients with the same index, which results in a system of linear equations in $t_m$:
\begin{align}
\label{LE_system}
\begin{split}
\left(\frac{1}{2} - \frac{1}{8}K_{00}\right)t_0 - \frac{1}{4}\sum_{n=1}^NK_{n0}t_n - \frac{1}{2}h_0 &= \frac{1}{2}f_0\\
\sum_{n=1}^N\left(\delta_{nm} - \frac{1}{2}K_{nm}\right)t_n - \frac{1}{4}K_{0m}t_0 - h_m &= f_m\ \ \ \ \mathrm{for}\ \ m=1,\ldots, N
\end{split}
\end{align}
Noting that $h_i$'s are linear combinations of $t_j$'s, the system of linear equations (equation~\ref{LE_system}) can be written in a concise matrix form as:
\begin{equation*}
Ft = f
\end{equation*}
which can be solved by standard matrix inversion techniques. The resulting solution ($t_m$) is used to calculate the temperature profile $\Delta\bar{T}\left(\eta, q, z\right)$ (equation~\ref{app:T_Fourier}) and the phonon energy distribution functions $G^+_\omega\left(z, \mu_i, \phi_j\right)$ and $G^-_\omega\left(z, -\mu_i, \phi_j\right)$ (equation~\ref{app:gen_soln_1}) for each $\eta$ and $q$ as follows:\\
First, the expressions for $I^+_{\mu\phi}$ and $I^-_{\mu\phi}$ are simplified as
\begin{align*}
I^+_{\mu\phi} &= C_\omega\int_0^d\Delta\bar{T}\exp\left(-\frac{\gamma^{\mathrm{FS}}_{\mu\phi}}{\mu\Lambda_\omega}z'\right)\mathrm{d}z' + \bar{Q}_\omega\tau_\omega\Lambda_\omega\frac{\mu}{\gamma^{\mathrm{FS}}_{\mu\phi}}\left(1-\exp\left(-\frac{\gamma^{\mathrm{FS}}_{\mu\phi}}{\mu\Lambda_\omega}d\right)\right)\\
&= C_\omega d\left(\frac{1}{2}t_0\frac{\mu\mathrm{Kn}^d_\omega}{\gamma^{\mathrm{FS}}_{\mu\phi}}\left(1-\exp\left(-\frac{\gamma^{\mathrm{FS}}_{\mu\phi}}{\mu\mathrm{Kn}^d_\omega}\right)\right) + \sum_{m=1}^Nt_mI_1\left(m\right)\right) \\
&\ \ \ \ \ + \bar{Q}_\omega\tau_\omega d\frac{\mu\mathrm{Kn}^d_\omega}{\gamma^{\mathrm{FS}}_{\mu\phi}}\left(1-\exp\left(-\frac{\gamma^{\mathrm{FS}}_{\mu\phi}}{\mu\mathrm{Kn}^d_\omega}\right)\right)\\
\end{align*}
\begin{align*}
I^-_{\mu\phi} &= C_\omega\int_0^d\Delta\bar{T}\exp\left(-\frac{\gamma^{\mathrm{FS}}_{\mu\phi}}{\mu\Lambda_\omega}z'\right)\mathrm{d}z' + \bar{Q}_\omega\tau_\omega\Lambda_\omega\frac{\mu}{\gamma^{\mathrm{FS}}_{\mu\phi}}\left(1-\exp\left(-\frac{\gamma^{\mathrm{FS}}_{\mu\phi}}{\mu\Lambda_\omega}d\right)\right)\\
&= C_\omega d\left(\frac{1}{2}t_0\frac{\mu\mathrm{Kn}^d_\omega}{\gamma^{\mathrm{FS}}_{\mu\phi}}\left(1-\exp\left(-\frac{\gamma^{\mathrm{FS}}_{\mu\phi}}{\mu\mathrm{Kn}^d_\omega}\right)\right) + \sum_{m=1}^Nt_mI_2\left(m\right)\right) \\
&\ \ \ \ \ + \bar{Q}_\omega\tau_\omega d\frac{\mu\mathrm{Kn}^d_\omega}{\gamma^{\mathrm{FS}}_{\mu\phi}}\left(1-\exp\left(-\frac{\gamma^{\mathrm{FS}}_{\mu\phi}}{\mu\mathrm{Kn}^d_\omega}\right)\right)\\
\end{align*}
where $\mathrm{Kn}^d_\omega = \Lambda_\omega/d$ is the Knudsen number of a phonon mode defined based on the thickness of the thin film.
Next, using the expressions for $I^+_{\mu\phi}$ and $I^+_{\mu\phi}$, the expressions for $\bar{c}^+_\omega\left(0, \mu_i', \phi_j'\right)$ and $\bar{c}^-_\omega\left(d, \mu_i', \phi_j'\right)$ are evaluated and finally, the expressions for $G^+_\omega\left(\hat{z}, \mu_i, \phi_j\right)$ and $G^-_\omega\left(\hat{z}, -\mu_i, \phi_j\right)$ are evaluated as,
\begin{align}
\label{g_soln}
\begin{split}
G^+_\omega\left(\hat{z}, \mu_i, \phi_j\right) &= \left(\sum_{i'j'}\left[T^{+}_{kk'}\bar{c}^+_\omega\left(0, \mu_{i'}, \phi_{j'}\right) + T^{-}_{kk'}\bar{c}^-_\omega\left(d, \mu_{i'}, \phi_{j'}\right)\right]\right)\exp\left(-\frac{\gamma^{\mathrm{FS}}_{ij}}{\mu_i\mathrm{Kn}^d_\omega}\hat{z}\right) \\
&\ \ \ \ \ +\ \frac{1}{4\pi\gamma^{\mathrm{FS}}_{ij}}\left(C_\omega\frac{t_0}{2} + \bar{Q}_\omega\tau_\omega\right)\left[1-\exp\left(-\frac{\gamma^{\mathrm{FS}}_{ij}}{\mu_i\mathrm{Kn}^d_\omega}\hat{z}\right)\right] + \frac{C_\omega}{4\pi\mu_i\mathrm{Kn}^d_\omega}\sum_{m=1}^Nt_mI^1_3\left(m; \hat{z}\right)\\
G^-_\omega\left(\hat{z}, -\mu_i, \phi_j\right) &= \left(\sum_{i'j'}\left[B^{+}_{kk'}\bar{c}^+_\omega\left(0, \mu_{i'}, \phi_{j'}\right) + B^{-}_{kk'}\bar{c}^-_\omega\left(d, \mu_{i'}, \phi_{j'}\right)\right]\right)\exp\left(-\frac{\gamma^{\mathrm{FS}}_{ij}}{\mu_i\mathrm{Kn}^d_\omega}\left(1-\hat{z}\right)\right) \\
&\ \ \ \ \ + \frac{1}{4\pi\gamma^{\mathrm{FS}}_{ij}}\left(C_\omega\frac{t_0}{2} + \bar{Q}_\omega\tau_\omega\right)\left[1 - \exp\left(-\frac{\gamma^{\mathrm{FS}}_{ij}}{\mu_i\mathrm{Kn}^d_\omega}\left[1-\hat{z}\right]\right)\right] + \frac{C_\omega}{4\pi\mu_i\mathrm{Kn}^d_\omega}\sum_{m=1}^Nt_mI^2_3\left(m; \hat{z}\right)\\
\end{split}
\end{align}
where,
\begin{align*}
\begin{split}
I^1_3\left(m; \hat{z}\right) &= \int_0^{\hat{z}'}\exp\left(-\frac{{\gamma^{\mathrm{FS}}_{ij}}}{\mu_i\mathrm{Kn}^d_\omega}\left[\hat{z}' - \hat{z}\right]\right)\cos\left(m\pi\hat{z}\right)\mathrm{d}\hat{z}\\
&= \frac{1}{m^2\pi^2 + \left(\frac{\gamma^{\mathrm{FS}}_{ij}}{\mu_i\mathrm{Kn}^d_\omega}\right)^2}\left[\left(\frac{\gamma^{\mathrm{FS}}_{ij}}{\mu_i\mathrm{Kn}^d_\omega}\cos\left(m\pi\hat{z}'\right) + m\pi\sin\left(m\pi\hat{z}'\right)\right) - \frac{\gamma^{\mathrm{FS}}_{ij}}{\mu_i\mathrm{Kn}^d_\omega}\exp\left(-\frac{{\gamma^{\mathrm{FS}}_{ij}}}{\mu_i\mathrm{Kn}^d_\omega}\hat{z}'\right)\right]\\
I^2_3\left(m; \hat{z}\right) &= \int_{\hat{z}'}^1\exp\left(\frac{{\gamma^{\mathrm{FS}}_{ij}}}{\mu_i\mathrm{Kn}^d_\omega}\left[\hat{z}' - \hat{z}\right]\right)\cos\left(m\pi\hat{z}\right)\mathrm{d}\hat{z}\\
&= \frac{1}{m^2\pi^2 + \left(\frac{\gamma^{\mathrm{FS}}_{ij}}{\mu_i\mathrm{Kn}^d_\omega}\right)^2}\Bigg[\left(-\frac{{\gamma^{\mathrm{FS}}_{ij}}}{\mu_i\mathrm{Kn}^d_\omega}\left(-1\right)^m\right)\exp\left(-\frac{{\gamma^{\mathrm{FS}}_{ij}}}{\mu_i\mathrm{Kn}^d_\omega}\left(1-\hat{z}'\right)\right) \\
&\ \ \ \ \ - \left(-\frac{\gamma^{\mathrm{FS}}_{ij}}{\mu_i\mathrm{Kn}^d_\omega}\cos\left(m\pi\hat{z}'\right) + m\pi\sin\left(m\pi\hat{z}'\right)\right)\Bigg]
\end{split}
\end{align*}
Once again, as in the steady state condition, these general solutions for $G_\omega$ are substituted into the expression for heat flux and the suppression in thermal conductivity due to phonon boundary scattering is derived for the transient transport condition in the main article.
\section{Monte Carlo Solution} ~\label{MC_BTE}
As a validation of our semi-analytical solution of the BTE, the time domain BTE (equation 1 in the main article) is also solved using a stochastic Monte Carlo technique to simulate thermal transport in transient grating experiment on a thin film. The Monte Carlo method is a particle based stochastic technique in which the computational particles representing phonon bundles are advected, scattered and sampled according to the governing time domain BTE. For this work, an efficient energy-based variance reduced formulation introduced by \textsl{Peraud et al.}~\cite{peraud_efficient_2011, peraud_alternative_2012} is used. For the steady state transport along a thin film, the simulation procedure is identical to the one described in ref.~\cite{peraud_efficient_2011}. For the transient simulation, the simulation domain consists of a pair of adiabatic (specularly reflecting) walls separated by an in-plane distance $L = \pi /q$, where $q$ is the grating wave vector and another pair of walls representing the cross-plane boundaries of the thin film with a separation $d$. Phonon bundles are initialized within the simulation domain with properties drawn according to an in-plane sinusoidal temperature profile as described in ref.~\cite{peraud_efficient_2011}. As the simulation evolves in time through advection and scattering of the phonon bundles with each other and with the boundaries, the temperature within a region as a function of time is obtained from the total time spent by each phonon bundle within that region. The resulting temperature profile as a function of time is fit to an exponentially decaying function whose decay rate is proportional to the thermal conductivity of the thin film at the grating wavevector $q$. \\

Three types of boundary scattering events (specular reflection, non-thermalizing diffuse scattering and thermalizing diffuse scattering) at the cross-plane walls of the thin film are implemented in our Monte Carlo simulations:
\begin{enumerate}
\item For the specular boundary condition, the outgoing phonon bundle retains the in-plane direction of propagation of the incoming phonon bundle while the cross-plane direction is reversed, that is,
\begin{align*} 
g^{\mathrm{out}}_\omega\left(z=0, \mu, \phi\right) &= g^{\mathrm{in}}_\omega\left(z=0, -\mu, \phi\right)\\
g^{\mathrm{out}}_\omega\left(z=d, \mu, \phi\right) &= g^{\mathrm{in}}_\omega\left(z=d, -\mu, \phi\right)\\
\end{align*}
Therefore, for every phonon bundle encountering the boundaries, heat flux is conserved.
\item For the non-thermalizing diffuse scattering, the incident phonon properties (frequency, polarization, group velocity and scattering time) are retained, while the new direction of propagation is sampled from $\int_{\mu=0}^1\int_{\phi=0}^{2\pi}\mu\mathrm{d}\mu\mathrm{d}\phi$. This procedure conserves heat flux at the boundary automatically, as described in ref.~\cite{peraud_monte_2014}.
\item For the thermalizing diffuse scattering, the phonons incident on the boundaries are terminated and new phonon properties are drawn from the distribution corresponding to $g^0_\omega\left(\Delta T\left(z=0/d\right)\right)/\tau_\omega$ which represents the equilibrating
 part of the BTE, while the new direction of propagation is sampled from $\int_{\mu=0}^1\int_{\phi=0}^{2\pi}\mu\mathrm{d}\mu\mathrm{d}\phi$. Although this procedure doesn't conserve heat flux at the boundary in general, the new phonons are drawn according to the boundary condition used in Fuchs-Sondheimer theory.
\end{enumerate}
\bibliographystyle{apsrev4-1}
%